\documentclass{ws-sd}

\begin{document}

\markboth{G. Litak, T. Kami\'nski, J. Czarnigowski, A.K. Sen, and M. Wendeker}{Combustion Process in a Spark Ignition
Engine}

%%%%%%%%%%%%%%%%%%% Publisher's Area please ignore %%%%%%%%%%%%%%%%%%%%%%%
\catchline{}{}{}{}{}
%%%%%%%%%%%%%%%%%%%%%%%%%%%%%%%%%%%%%%%%%%%%%%%%%%%%%%%%%%%%%%%%%%%%%%%%%%

\title{COMBUSTION PROCESS IN A SPARK IGNITION ENGINE: ANALYSIS OF CYCLIC MAXIMUM PRESSURE AND PEAK  PRESSURE ANGLE}

\author{GRZEGORZ LITAK}
\address{Department of Applied Mechanics, Technical University of
Lublin, Nadbystrzycka 36, PL-20-618 Lublin, Poland \\
and \\
 Department of Mechanical Engineering,
        University of Trieste,
Via A. Valerio 10, I-34127, Trieste, Italy \\
g.litak@pollub.pl
}

\author{TOMASZ KAMI\'NSKI}

\address{Motor Transport Institute,
ul. Jagiello\'nska 80, PL-03-301
Warsaw, Poland \\
tk42@interia.pl}

\author{JACEK
CZARNIGOWSKI}

\address{Department of Machine Construction, Technical University of
Lublin, Nadbystrzycka 36, PL-20-618 Lublin, Poland \\
j.czarnigowski@pollub.pl}

\author{ASOK K. SEN}

\address{Department of Mathematical Sciences, Indiana University,  
402 N. Blackford Street,
Indianapolis, IN 46202-3216 \\
asen@iupui.edu}

\author{MIROS\L{}AW  
WENDEKER}

\address{Department of Combustion Engines, Technical University
of
Lublin,
Nadbystrzycka 36, PL-20-618 Lublin, Poland \\
m.wendeker@pollub.pl}

\maketitle

\begin{history}
\received{(16 November 2006)}
%\received{(Day Month Year)}
%\revised{(Day Month Year)}
\end{history}

\begin{abstract}
 In this paper we analyze the cycle-to-cycle variations of maximum pressure $p_{max}$ and peak pressure angle $\alpha_{pmax}$ in a 
 four-cylinder spark ignition engine. We examine the experimental time series of $p_{max}$ and $\alpha_{pmax}$ for three different spark 
advance angles. Using standard statistical techniques such as return maps 
 and histograms we show that depending on the spark advance angle, there are significant differences in the fluctuations of $p_{max}$ and 
 $\alpha_{pmax}$. We also calculate the multiscale entropy of the various time series to estimate the effect of randomness in these 
 fluctuations. Finally, we explain how the information on both  $p_{max}$ and $\alpha_{pmax}$ can be used to develop optimal strategies for 
controlling the combustion process and improving engine performance.
\end{abstract}

\keywords
{stochastic process;  stochastic analysis; nonlinear time series analysis; combustion process}

\ccode{AMS Subject Classification: 60G35, 60H99, 37M10}
%60G35- Stochastic processes,Applications;
%60H99- Stochastic analysis;
%37M10- Approximation methods and numerical treatment of dynamical systems, Time series analysis}

\section{Introduction}

Instabilities in combustion processes were observed
from the very beginning of the Spark Ignition (SI) engine development
\cite{Clerk1886}. These instabilities may cause 
fluctuations in the  power
output as well as fluctuations of the burned fuel mass.
As a consequence, the mean effective torque may be reduced by as much as 20$\%$.
It is therefore not surprising that the instabilities were  identified as 
a fundamental 
combustion
problem in spark ignition engines \cite{Patterson1966}.
A disturbing feature of these instabilities is 
the unpredictable character of their occurrence, which
makes an engine
difficult to control 
\cite{Hubbard1976,Matekunas1986,Sawamoto1987,Wagner1993,Eriksson1997}. 
Recognition and elimination of their sources have been one of the main issues in SI engine technology 
in the last century 
\cite{Heywood1988,Hu1996,Wagner2001,Daw2003}.  
However, in spite of great efforts made in clarifying the various aspects of 
combustion instabilities in the past several years, 
the problem of a stable combustion process control in an SI engine has 
not yet been
solved.\cite{Wendeker2003,Kaminski2004}. The main factors of combustion
instabilities as classified by Heywood \cite{Heywood1988} are aerodynamics in the cylinder
during combustion, amounts of fuel, air and recycled exhaust gases supplied
to  the cylinder, and composition of the local mixture near the spark plug.
Among the various researchers, Winsor {\em et al.} \cite{Winsor1973} explored the turbulent aspects of 
combustion and studied the nature of pressure fluctuations in an SI engine. Recent attempts have  
focused on the development of nonlinear dynamical models of the combustion process. For example, 
Daily \cite{Daily1988} formulated a simple nonlinear model and demonstrated that pressure 
variations of a chaotic type can originate from the initial conditions prevailing
at the beginning of each cycle. Using a nonlinear model, Kantor \cite{Kantor1984} analyzed the 
cycle-to-cycle variations of the process variables including combustion temperature. Foakes and 
Pollard \cite{Foakes1993}, Chew {\em et al.} \cite{Chew1994} and Letellier {\em et al.}  \cite{Letellier1997} 
examined nonlinear models of pressure variations and calculated their Lyapunov exponents \cite{Letellier1997}.
Subsequently, Daw {\em et al.} \cite{Daw1996,Daw1998} included the effect of exhaust gas
circulation in a nonlinear model to investigate engine dynamics. But the most convincing argument for the presence of nonlinear 
dynamics in the combustion process came from the concept of time irreversibility of heat release as shown 
by Green {\it et al.} \cite{Green1999}, Wagner {\it et al.}
\cite{Wagner2001} and Daw {\it et al.} \cite{Daw2003}. More recently, on the basis of their analysis of 
cycle-to-cycle pressure variations, Wendeker  {\it et al.} \cite{Wendeker2004} proposed an 
intermittency mechanism leading to chaotic combustion.

In this paper we continue these studies with an examination of the experimental time series of maximum 
pressure $p_{max}$ and peak pressure angle $\alpha_{pmax}$ for three different spark advance angles.
Our paper is divided into four sections. After this introductory section (Sec. 1), we provide a 
description of our experimental standing in Sec. 2 and outline the procedure of 
performing measurements. This is followed by an analysis of the  statistical properties of the time series
of $p_{max}$ and $\alpha_{pmax}$ using return maps and histograms. We also calculate the multiscale entropy 
(MSE) of the various time series in order to estimate their complexity and delineate the effect of randomness in the fluctuations of $p_{max}$ and $\alpha_{pmax}$. Finally, in Sec. 4, we discuss the results of our analysis 
and use them to propose a criterion for optimal efficiency of combustion.

\section{Experimental Stand for Pressure Measurement}

%f1
\begin{figure} 
\vspace*{0.0cm}
\centerline{\includegraphics[width=12cm,angle=0]{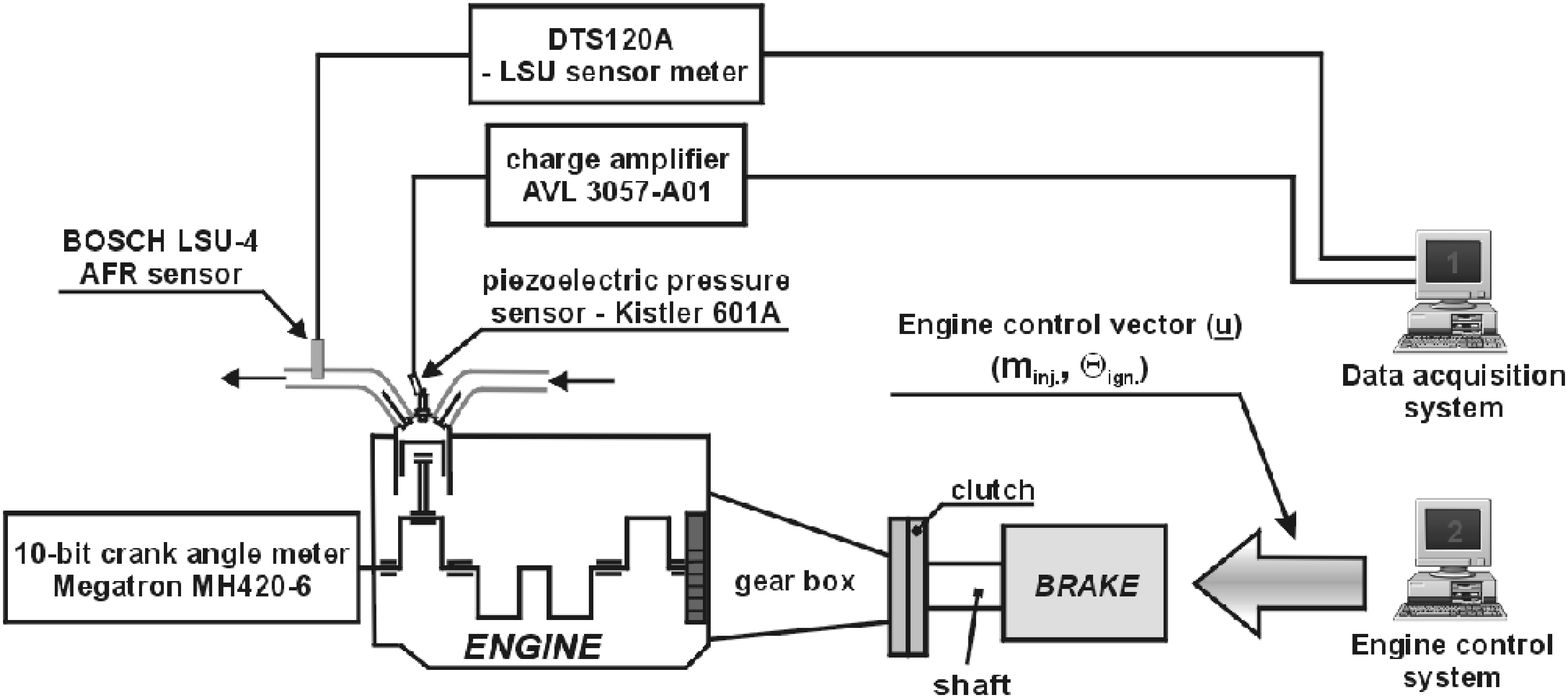}}
\caption{\label{Fig_one}
Schematic diagram of the experimental stand for pressure measurements.
}
\end{figure}

In our experimental stand (Fig. \ref{Fig_one}), the pressure inside the cylinder is measured
by a standard method  using a piezo-electric sensor.
It should be noted that pressure is the best known quantity that can be directly 
measured to analyze engine dynamics 
\cite{Sawamoto1987,Heywood1988,Matekunas1983,Ozdor1994,Taccani2003,Litak2005}.
Actual cylinder pressure together with the internal cylinder  volume 
 can be used to
obtain the indicated mean effective
 pressure (IMEP), to  calculate the engine torque and efficiency, and also to estimate the 
magnitudes of important process variables such as burning
rate, bulk temperature and heat release.
Furthermore, statistical analysis of internal pressure data can provide information 
about the stability of the combustion process.

Our equipment provides a direct measure of the pressure inside the cylinder of a spark ignition engine.
Internal pressure data were recorded at the Engine
Laboratory of the Technical University of Lublin, where we conducted a series of
tests. The
pressure traces were generated on a
 1998 cm$^3$ Holden 2.0 MPFI engine running at 1000 revolutions per minute. 
Measurements were made with a resolution of about $0.7^o$ of 
crankshaft revolution.
A single experiment consisted of about 2000 cycles of engine work.
Data were collected  at different spark timings with spark advance angles of 5, 15 and 30
degrees before the top dead center (TDC). The engine
speed, fuel-air ratio, and throttle setting were all held constant throughout the data 
collection period. Intake air pressure was maintained at 40 kPa and a stoichiometric mixture was used.

\begin{figure}

{\center \includegraphics[width=11cm,angle=0]{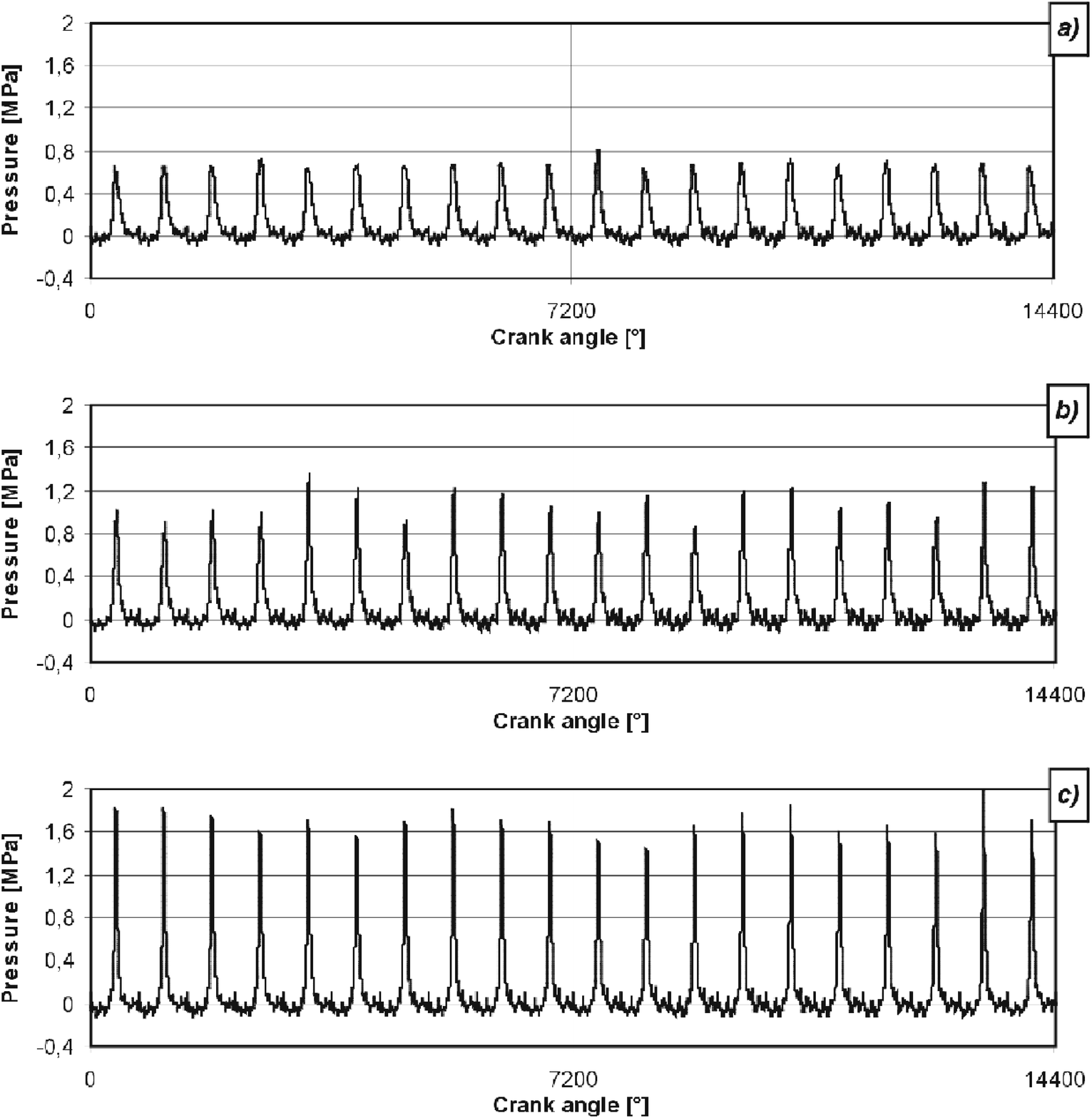}}
\caption{
\label{Fig_two}
Variations of internal pressure with crank angle during the first 20 combustion cycles.
The panels (a), (b) and (c) correspond to the spark advance angles
$\Delta \alpha_z=5^o$, 15$^o$ and 30$^o$, respectively.}
\end{figure}

A few examples of the pressure measurements are presented in Fig. \ref{Fig_two}. Here we have 
plotted the variation of pressure with crank angle during the first 20 combustion cycles for three  
different values of the spark advance angles: $\Delta \alpha_z=5^o$, 15$^o$ and 30$^o$. Note that 
the pressure measurements reflect 
the combined effects of cylinder volume compression and combustion of the fuel-air mixture.
From these measurements the values of maximum pressure $p_{max}$ and the peak 
pressure angle $\alpha_{pmax}$ can be easily identified.
 
%f2

Before we go into a detailed 
analysis of the time series of $p_{max}$ and $\alpha_{pmax}$ in 
consecutive combustion cycles, we would 
like to point out  the importance of these quantities for combustion control and diagnostics. 
Note that the  output torque usually scales    
as the square of the mean values of $p_{max}$ or $\alpha_{pmax}$ for 
different speeds \cite{Eriksson1997,Nielsen1998,Kaminski2005}.

\section{Cyclic Time Series and Pressure Fluctuations}

%f3
\begin{figure}
\vspace*{0.0cm}
\centerline{\includegraphics[width=12cm,angle=0]{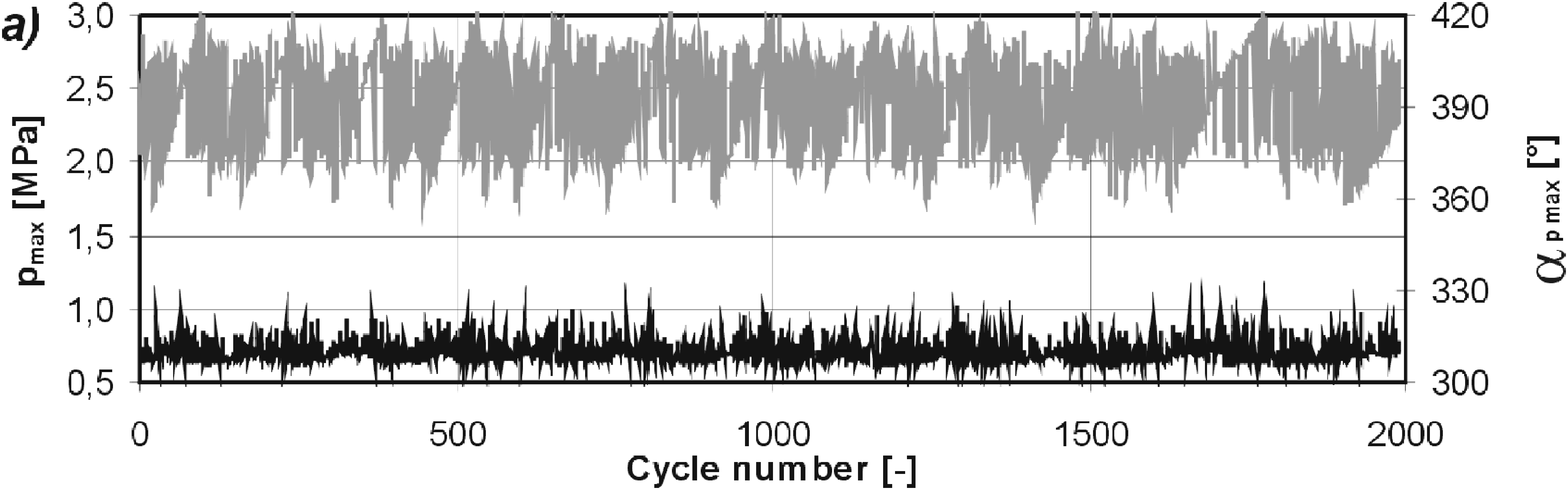}}
\centerline{\includegraphics[width=12cm,angle=0]{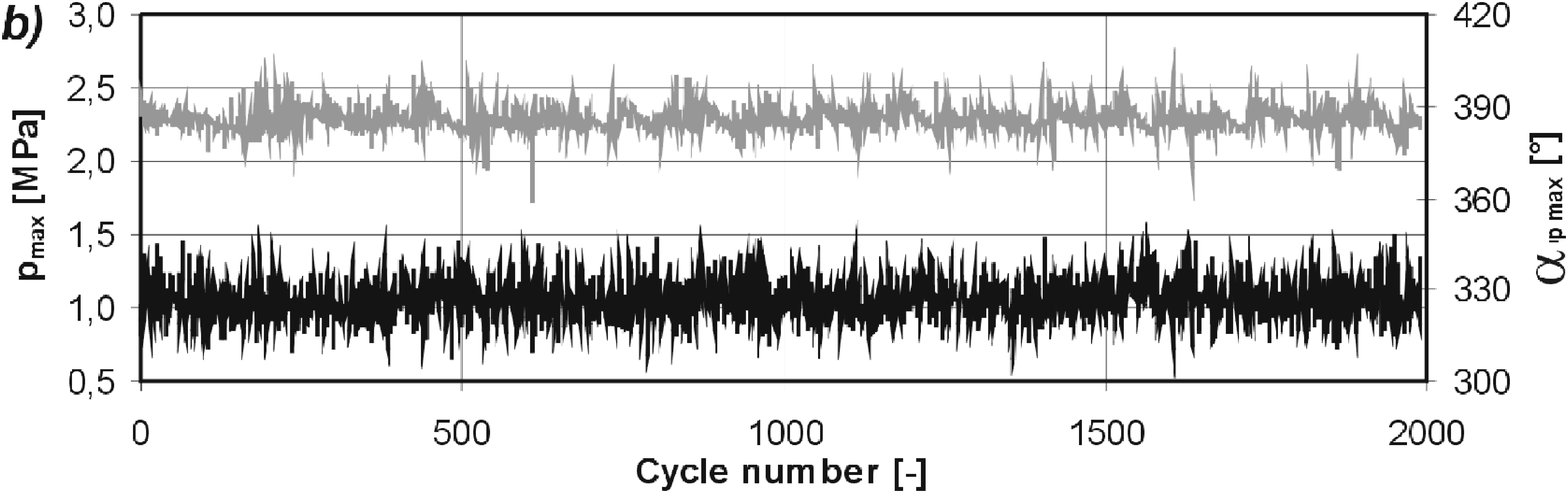}}
\centerline{\includegraphics[width=12cm,angle=0]{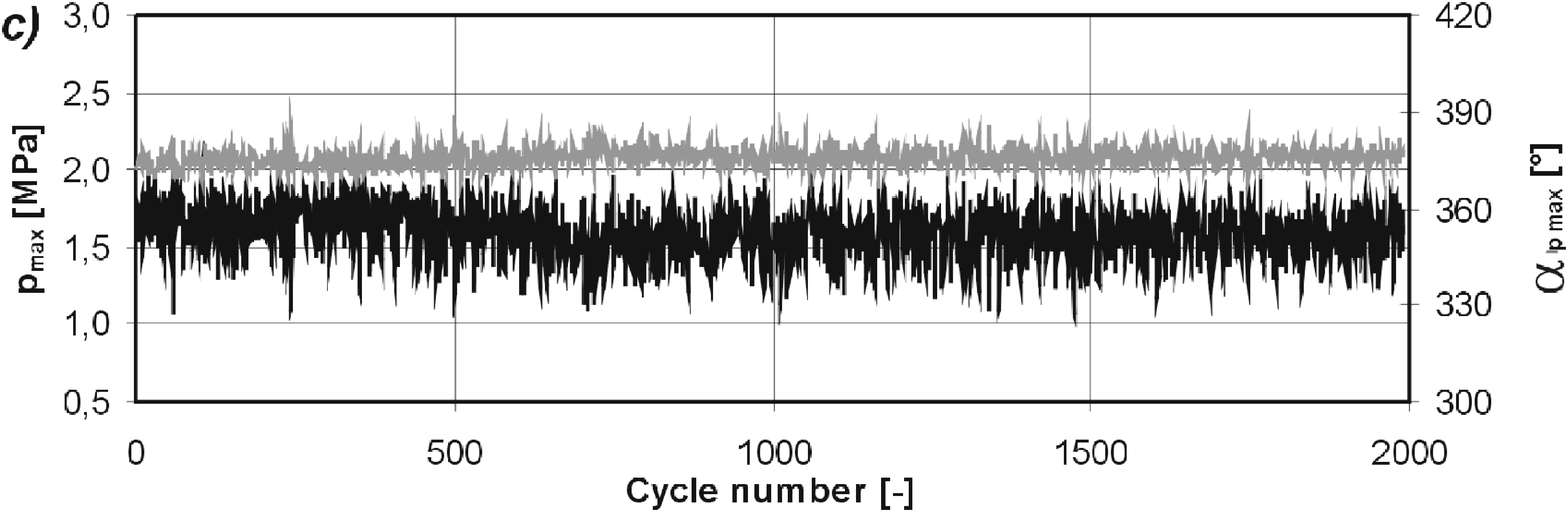}}
\caption{\label{Fig_three}
Time series of cycle-to-cycle variations of  maximum pressure  $p_{max}$ (in black)
and $\alpha_{pmax}$ (in grey) for three values of the spark advance angle:
$\Delta \alpha_z=5^o$, 15$^o$ and 30$^o$, for Figs. \ref{Fig_three}a-c, respectively.
}
\end{figure}

From Fig. \ref{Fig_two}  we observe that  the cycle-to-cycle fluctuations of pressure are 
present in all cases considered and  
they change with the spark advance angle $\Delta \alpha_z$. It 
should be  noted 
that 
the magnitude of 
peak pressure, which is directly related to the power output, 
depends 
on  $\Delta \alpha_z$; $p_{max}$ is higher for larger $\Delta \alpha_z$. 
Unfortunately, the advantage of maximum pressure increase is offset 
 by  the disadvantage of increasing
fluctuations. In terms of the peak pressure $p_{max}$, 
this effect  can be observed in Figs. \ref{Fig_three}a-c.
Here we have also plotted the time series of the corresponding angles
$\alpha_{pmax}$ for which the peak pressure $p_{max}$ is reached.
Surprisingly, changes in combustion conditions which lead to increased  
fluctuations of $p_{max}$
are associated with decreased fluctuations of $\alpha_{pmax}$.

\subsection{Return Maps}

\begin{figure}
\vspace*{0.0cm}
\centerline{
\includegraphics[width=6.0cm,angle=0]{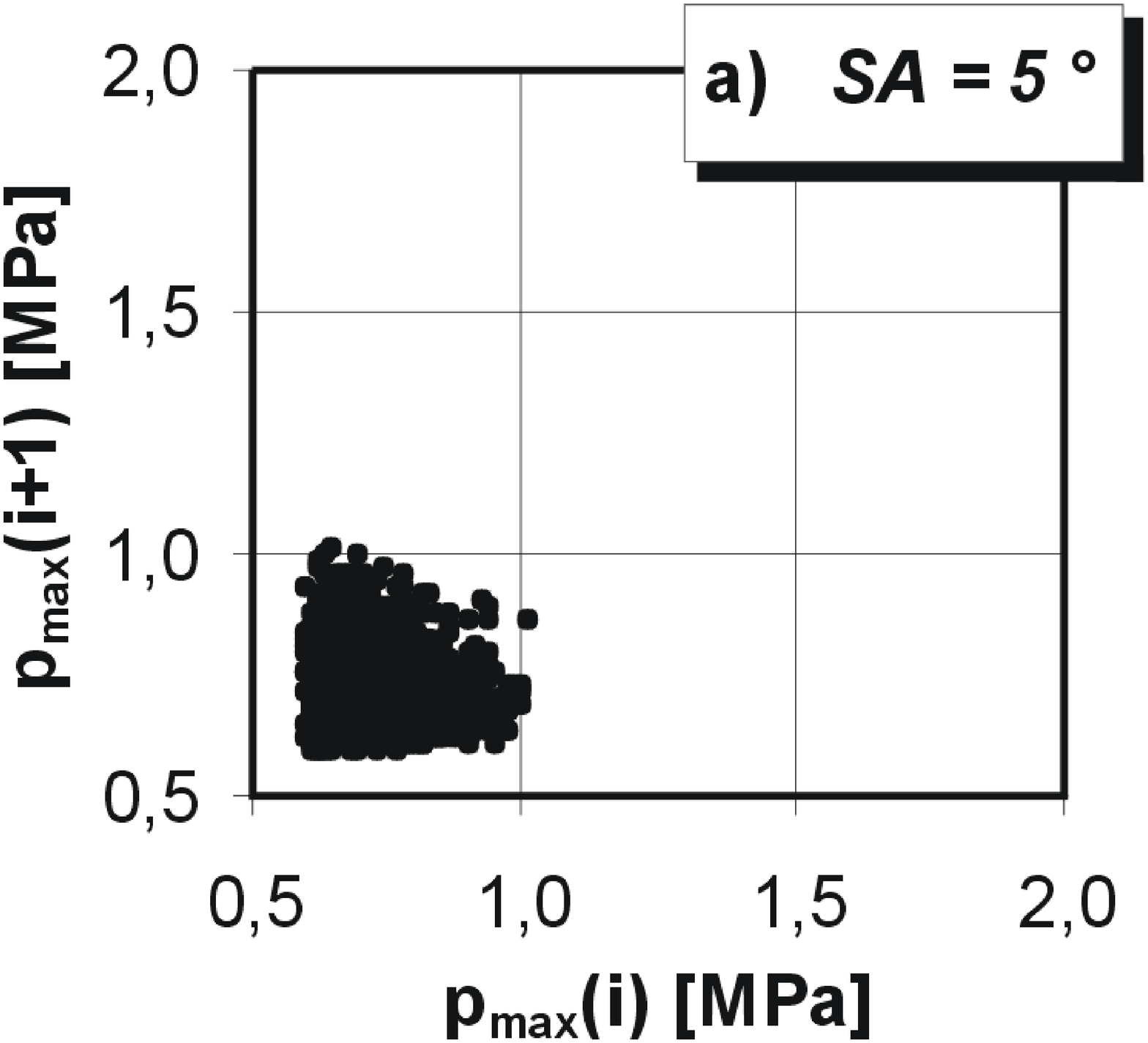}
\includegraphics[width=6.0cm,angle=0]{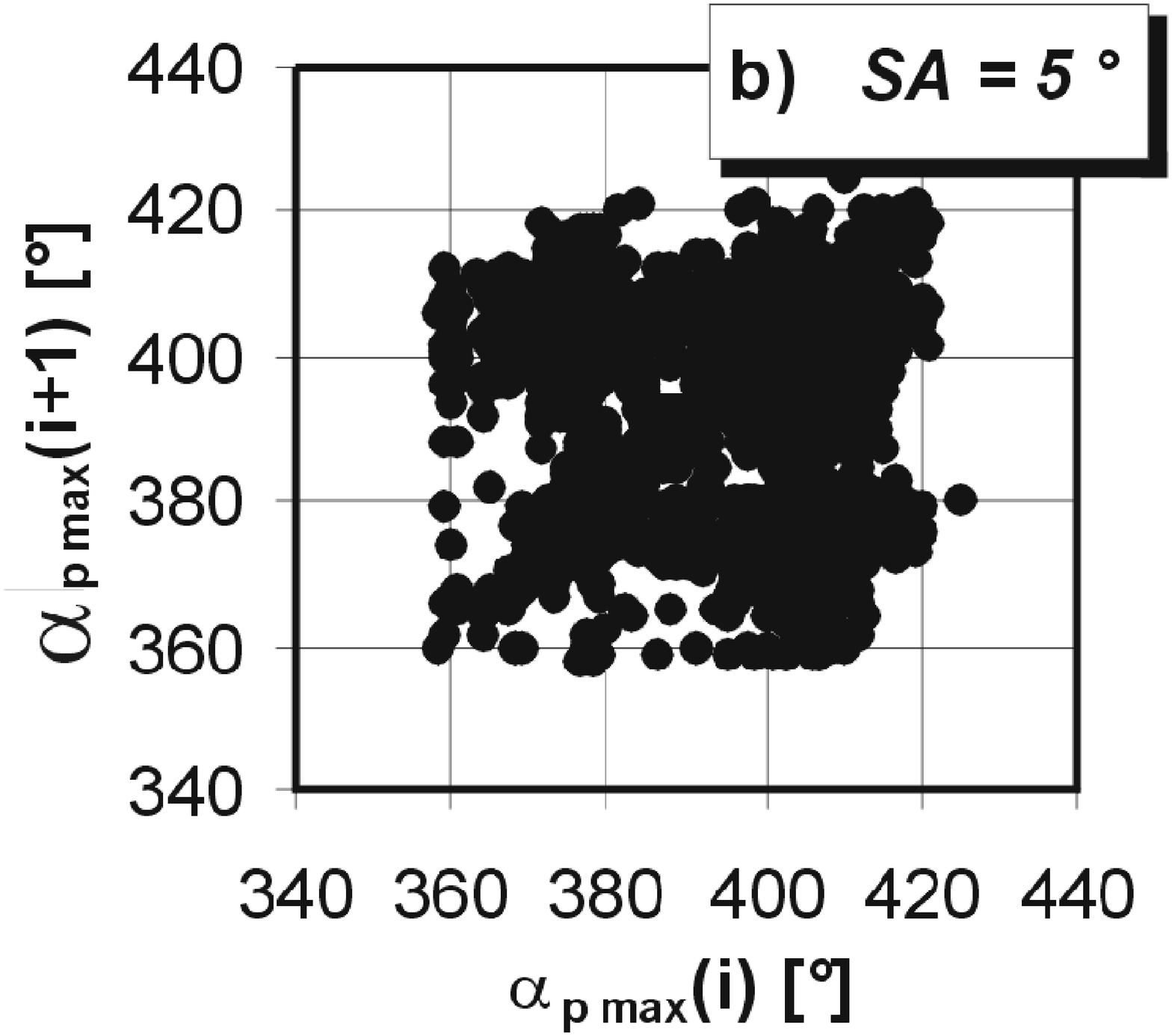}}
\centerline{\
\includegraphics[width=6.0cm,angle=0]{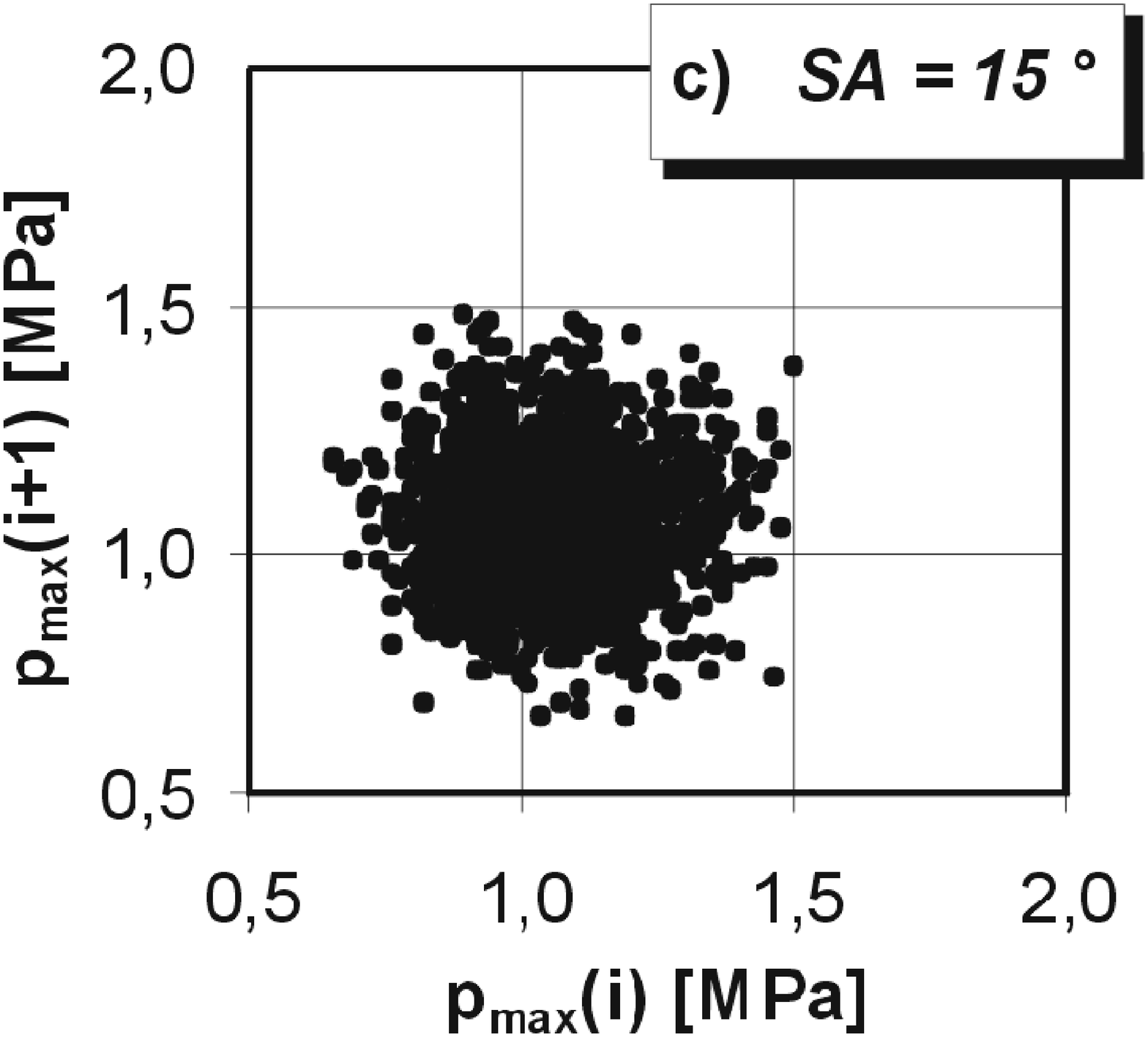}
\includegraphics[width=6.0cm,angle=0]{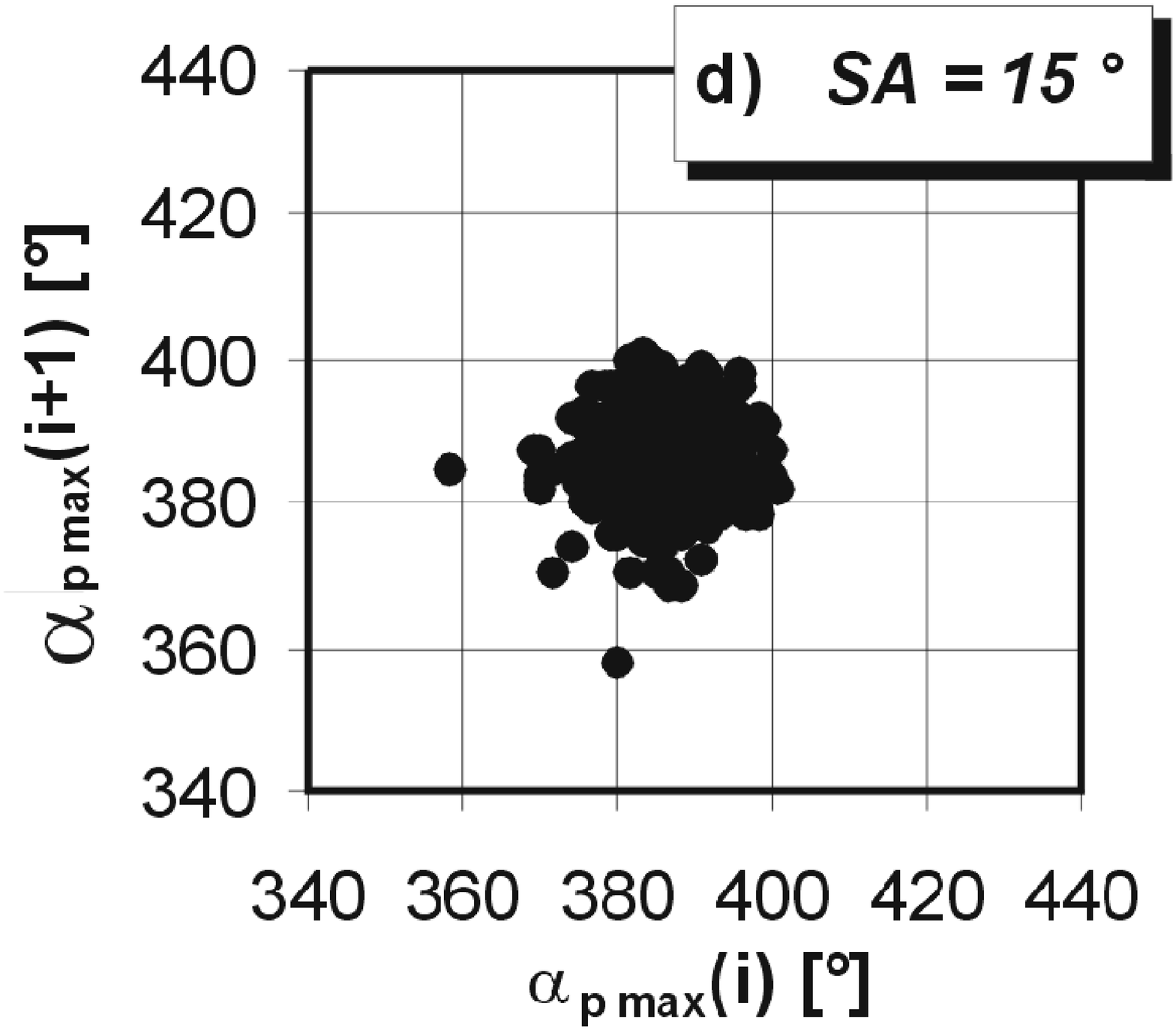}}
\centerline{\
\includegraphics[width=6.0cm,angle=0]{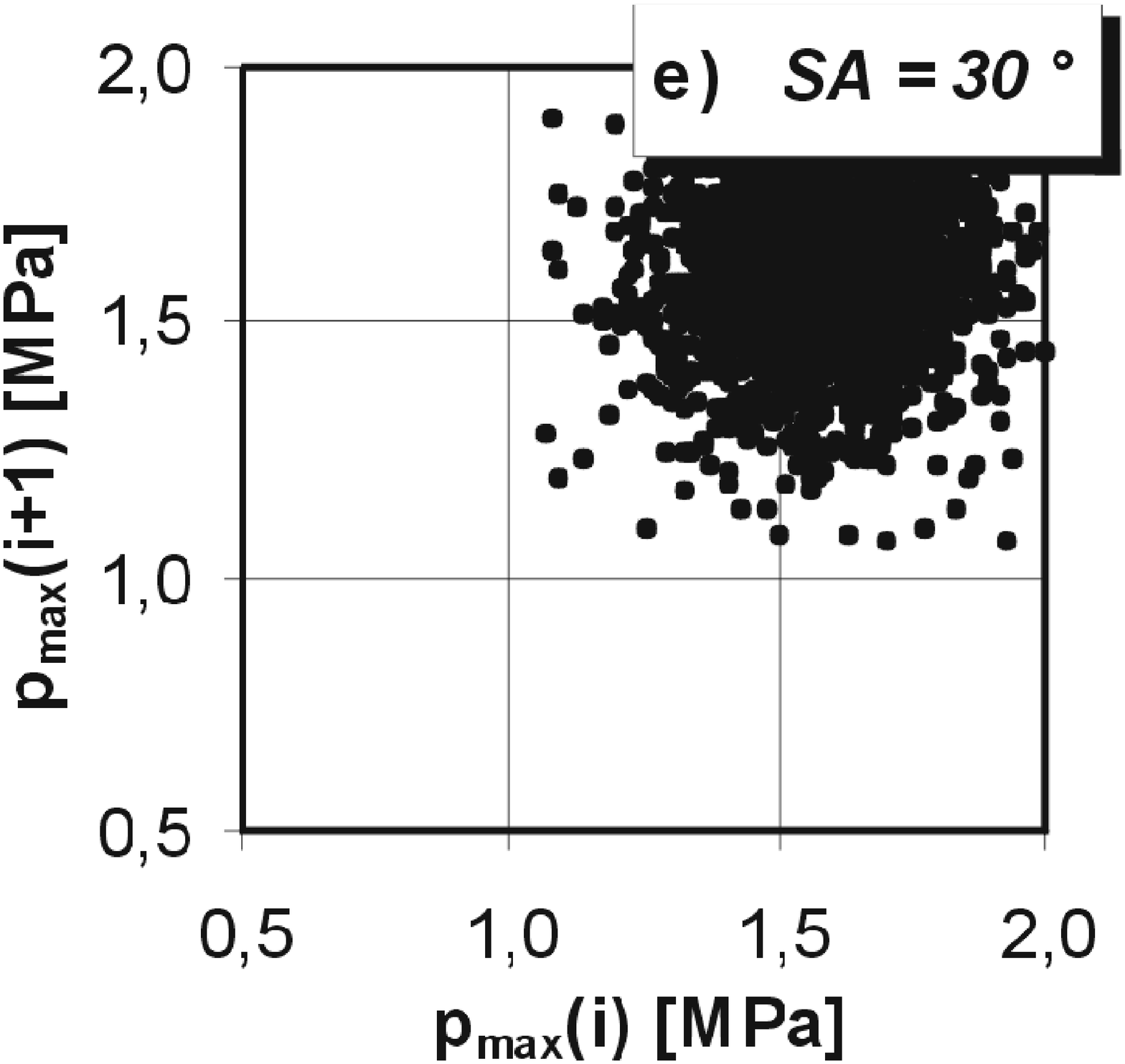}
\includegraphics[width=6.0cm,angle=0]{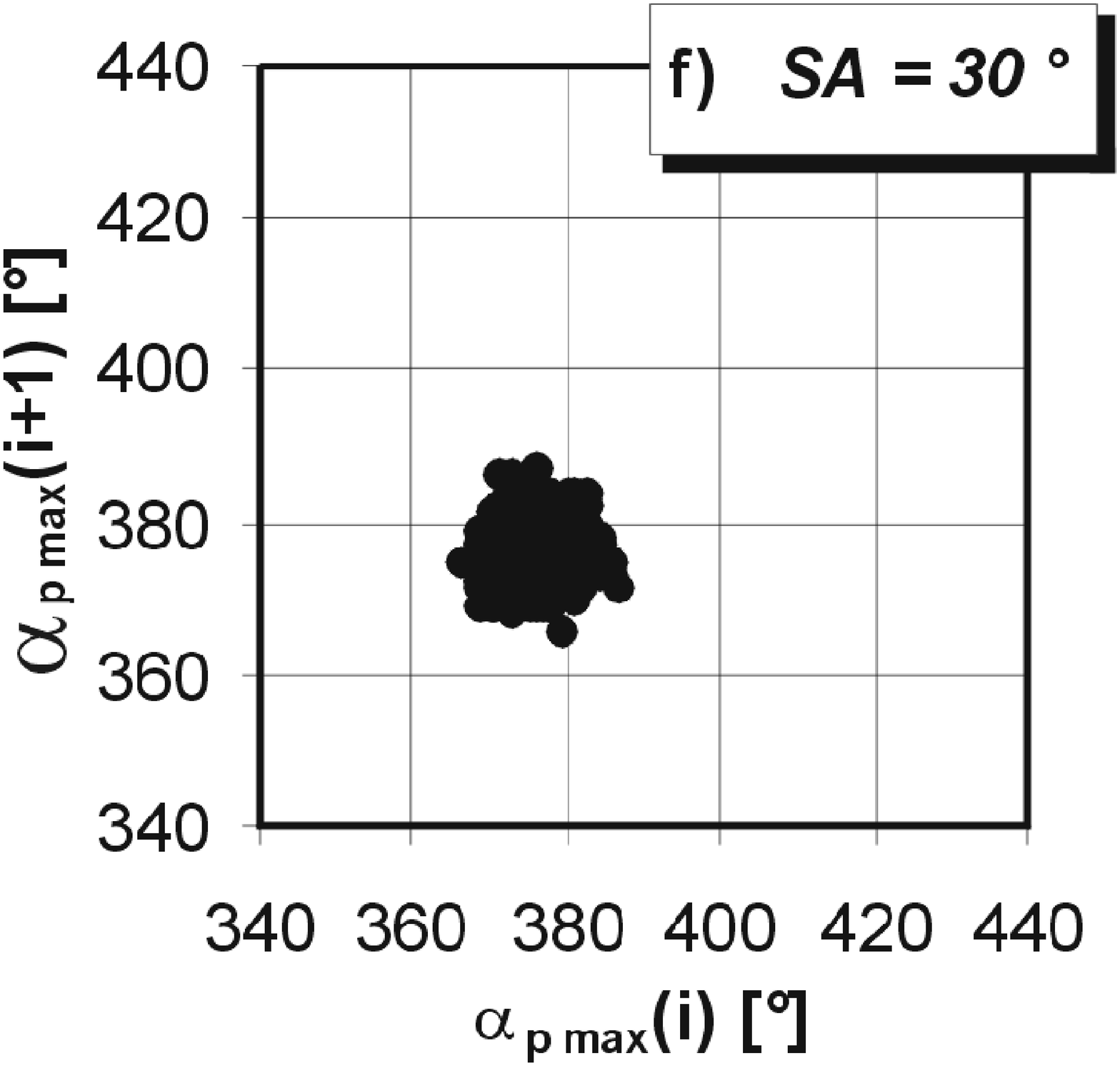}}

\caption{\label{Fig_four}
Return maps  of cyclic pressure maxima $p_{max}$ and  the corresponding
angles
$\alpha_{pmax}$
for three
spark advance angles
$\Delta \alpha_z$ denoted by SA$=5^o$, 15$^o$ and 30$^o$.
}
\end{figure}

To explore the properties of the time series of $p_{max}$ and $\alpha_{pmax}$ in more detail,
 we have plotted their return maps for 
the various spark advance angles. These are shown in Fig. \ref{Fig_four}. 
We define a sequential transformation for any cycle $i$:
\begin{equation}
F(i) \rightarrow F(i+1),
\end{equation}

%f4

\noindent
where $F$ can be  identified with $p_{max}$ or  $\alpha_{pmax}$. In a 
return map, black points in the coordinates  ($F(i+1)$,$F(i)$)  are marked.   
Such maps are very useful in examining 
changes of the corresponding quantity $F(i)$.  
The method of return maps has become a standard tool in 
the investigation of engine dynamics 
 but they are usually applied to burned mass ratio or heat release per cycle
\cite{Daw1996,Green1999,Litak2005a}. 
Let us analyze the first diagram (Fig. \ref{Fig_four}a) which corresponds
to the pressure history in Fig. \ref{Fig_two}a ($\Delta \alpha_z=5^o$). 
The characteristic black right angle in the left bottom
corner of the return map indicates a very low level of combustion or  misfire instabilities. 
The corner point represents the lowest value of $p_{max}$ which can be identified as 
peak compression  pressure.
In terms of peak pressure angles $\alpha_{pmax}$ (Fig. \ref{Fig_four}b) we observe that the 
points lying at the lowest angular position around 360$^o$ (for $\Delta \alpha_z=5^o$)
correspond to a maximum compression point -- the Top Dead Point (TPD). 
However, in this case ($\Delta \alpha_z=5^o$) we cannot draw any conclusion about misfires
because  we are not analyzing heat release here.
Extended studies of this effect can be found in   \cite{Green1999,Wagner2001,Litak2005a,Piernikarski2000}. 
In  Figs.  \ref{Fig_four}c,d and Figs.  \ref{Fig_four}e,f, we show the return maps for the spark
advance angles  $\Delta \alpha_z= 15^o$ and $30^o$, respectively.
In these cases the fluctuations may have a different origin. 
Following the previous analysis for $\Delta \alpha_z=15^o$  (Figs.  \ref{Fig_four}c,d),
we have detected  a singular point
with  relatively weak combustion, while for $\Delta \alpha_z=30^o$  
such points have disappeared. This means that the identification of pressure peak caused by combustion is  certain
in that regime.
Of course, the measured fluctuations will change with changes in the spark advance angle  $\Delta 
\alpha_z$ \cite{Litak2005}.
Statistical features covered by the return maps are in 
agreement with the
 time series histories depicted in Fig. \ref{Fig_three}a-c. Note that the fluctuations in $p_{max}$ increase while those in $\alpha_{pmax}$ 
decrease with increasing $\Delta
\alpha_z$.
We should also mention that all the maps possesses the approximate diagonal 
($F(i+1)$ versus $F(i)$) symmetry; this can
be interpreted as the time reversal symmetry of the corresponding time series of $p_{max}$ and $\alpha_{pmax}$.  
This symmetry has  not been broken in contrast to other reports on combustion experiments  
(Green {\em at al.} \cite{Green1999}). 
The main difference is that Green {\em at al.} \cite{Green1999} examined  the lean combustion limit 
whereas our fuel-air mixture is stoichiometric.
Clearly, preserved time-reversal symmetry  may 
appear if the stochastic 
component is 
significant. Alternatively, for deterministic but
chaotic time series one usually observes asymmetry in the return maps \cite{Green1999} associated with
a broken time-reversal symmetry.

\subsection{Histograms}

%f5
\begin{figure}
\centerline{\
\includegraphics[width=6.5cm,angle=0]{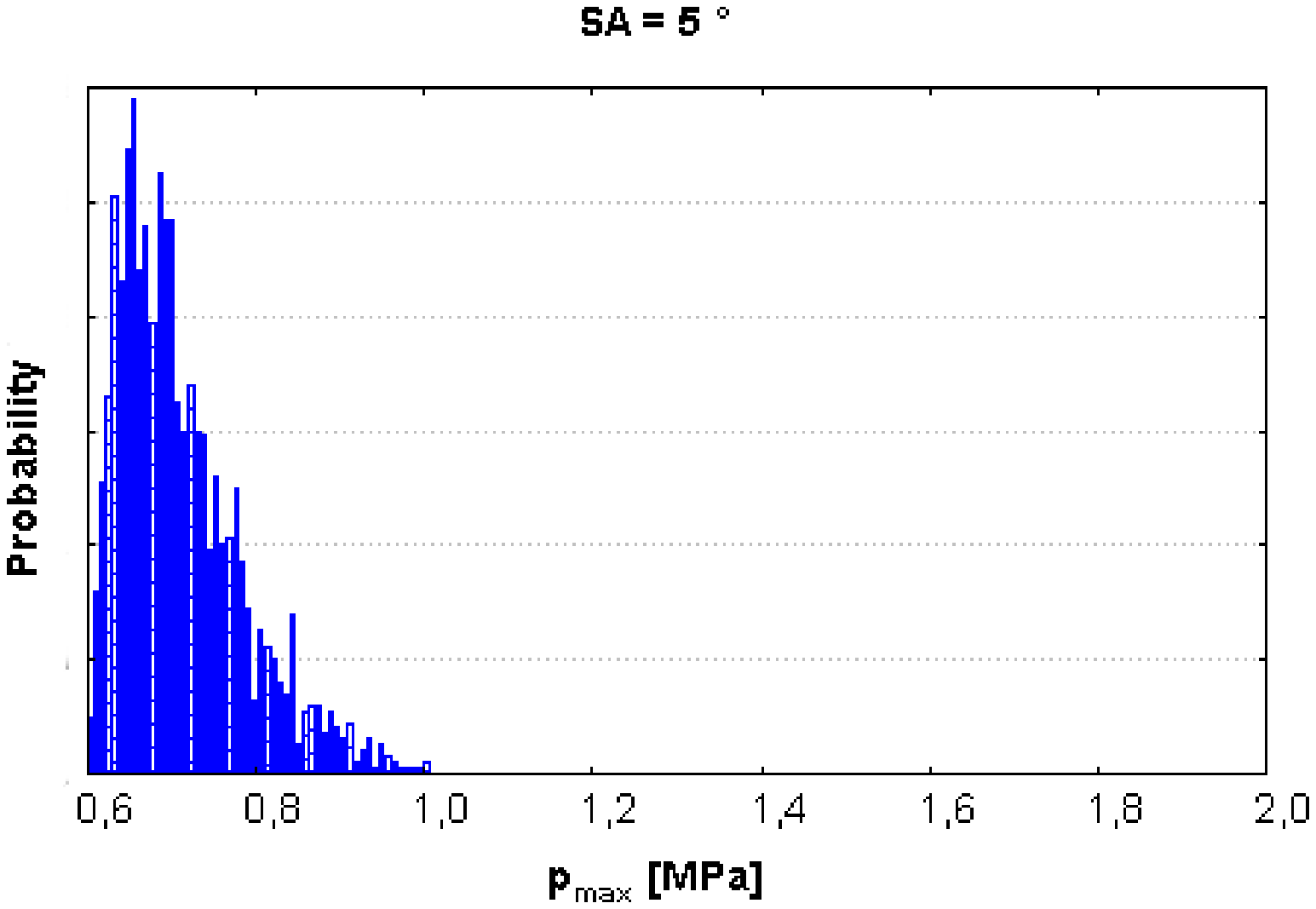} 
\includegraphics[width=6.5cm,angle=0]{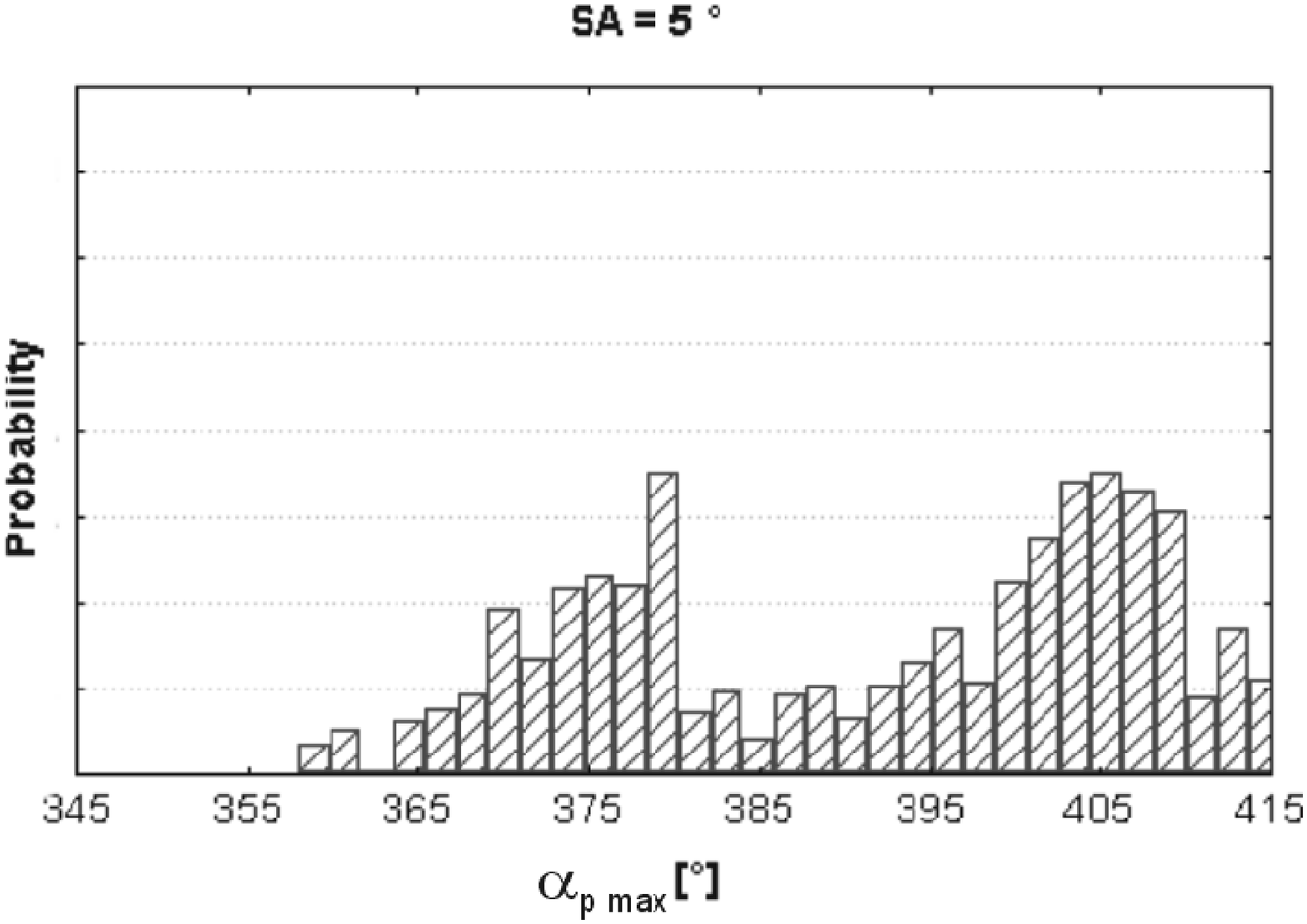}}

\hspace{0.5cm} (a) \hspace{6cm} (b)

\centerline{\
\includegraphics[width=6.5cm,angle=0]{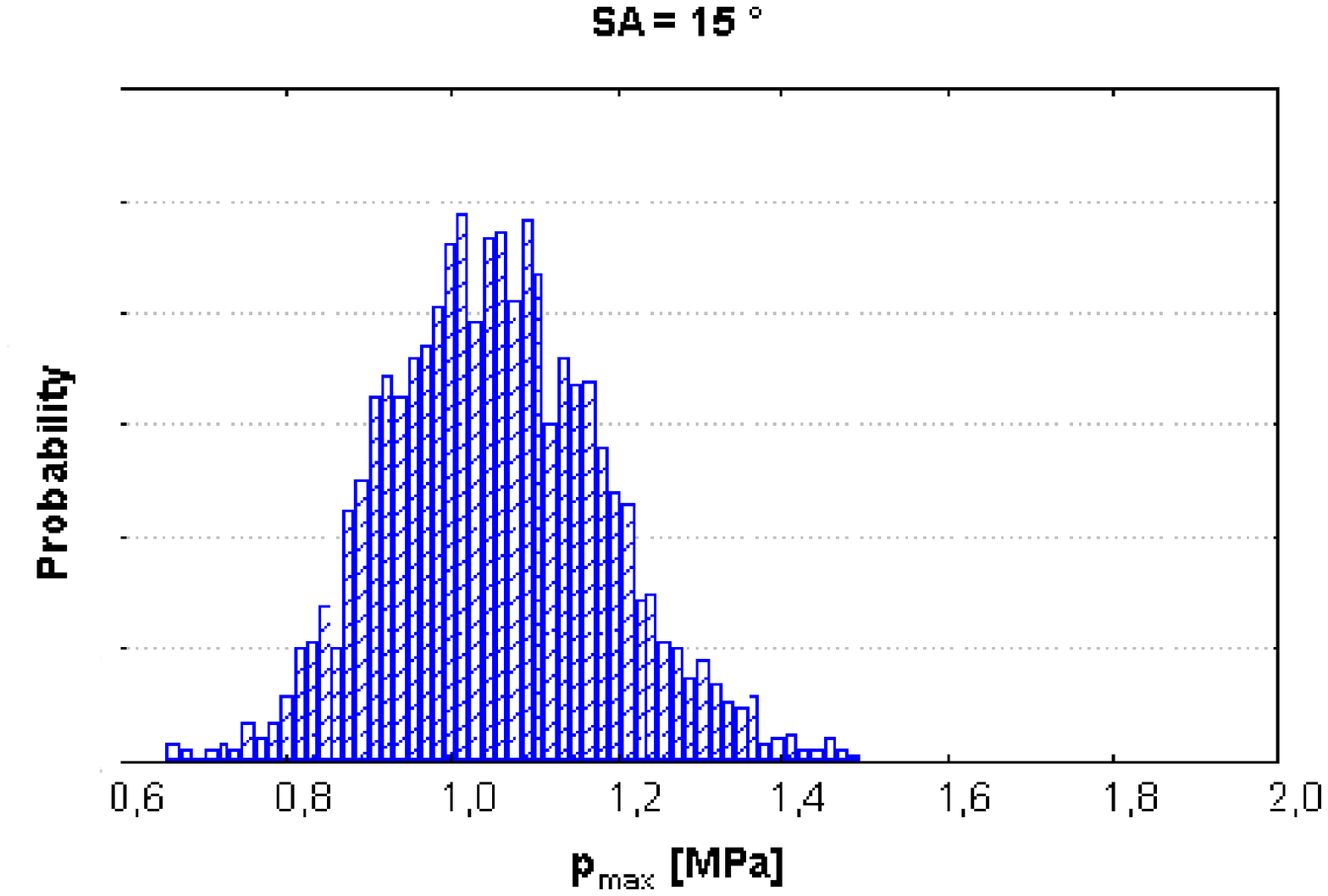}
\includegraphics[width=6.5cm,angle=0]{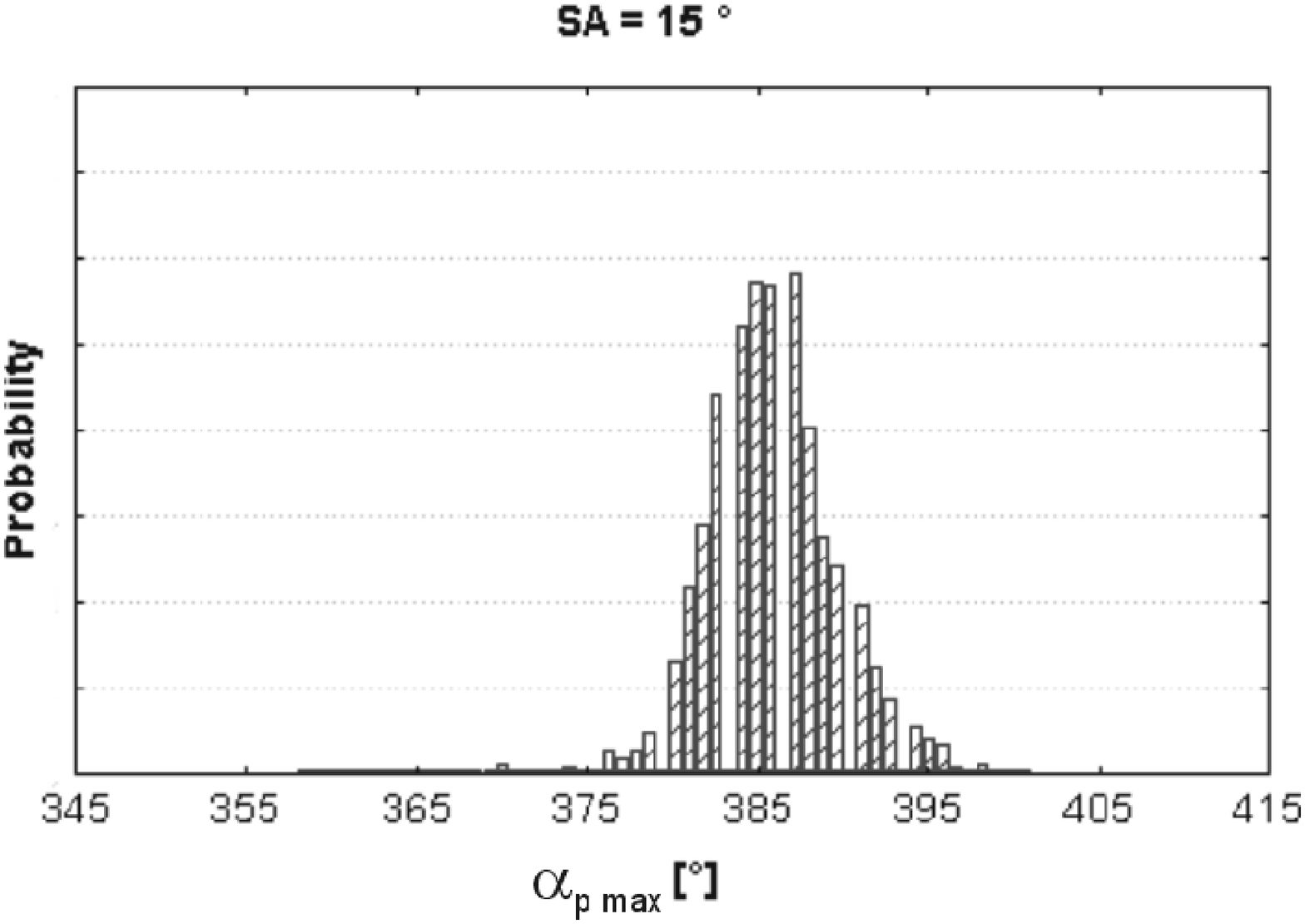}} 

\hspace{0.5cm} (c) \hspace{6cm} (d)

\centerline{\
\includegraphics[width=6.5cm,angle=0]{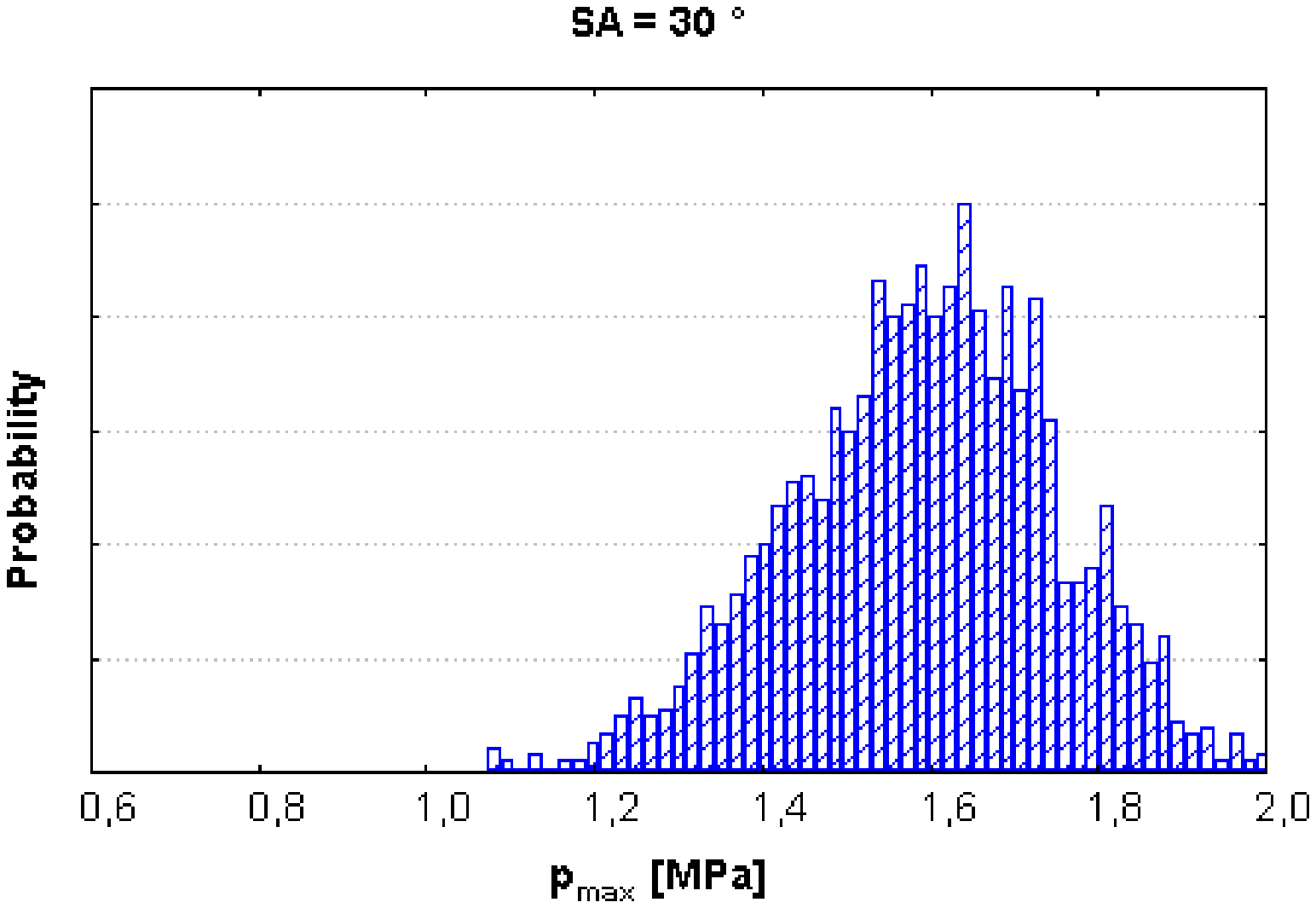}
\includegraphics[width=6.5cm,angle=0]{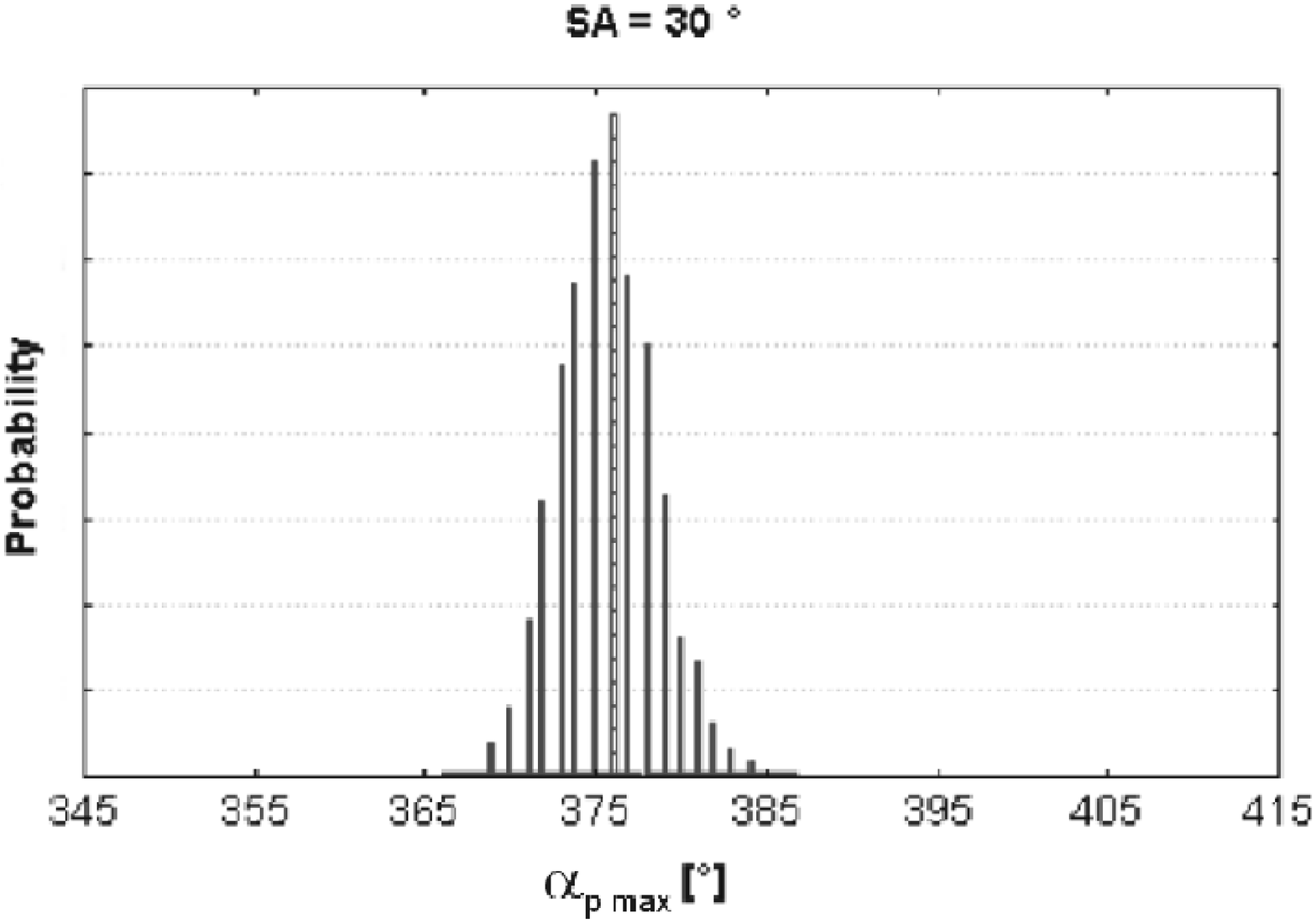}}

\hspace{0.5cm} (e) \hspace{6cm} (f)

\caption{\label{Fig_five}
Histograms of cyclic pressure maxima $p_{max}$ and the corresponding angles
 $\alpha_{pmax}$ ($SA=\Delta \alpha_z=5^o,~15^o,~30^o$).
}
\end{figure}

Here we perform further analysis of the $p_{max}$ and  $\alpha_{pmax}$ time series 
using
histograms.
From the time series we calculate the probability $P_n$: 
\begin{equation}
P_n= \frac{1}{N_{total}}\sum_i^{N_{total}} \Theta(\tilde F_n-F(i)) \Theta(F(i) 
-\tilde F_{n+1}), 
\end{equation} 
where $\Theta(x)$ is the Heaviside step function, $F(i)$ is a sequentially 
measured  quantity, and $\tilde F_n$ denotes a  discretized value.  
Here $N_{total}=1991$ is the total number of 
combustion cycles in a single  time series. 
In Figs. \ref{Fig_five}a-f  we have plotted the histograms of the time series for $p_{max}$ or 
$\alpha_{pmax}$ for the various cases considered in Figs. \ref{Fig_four}a-f. 
Interestingly, the combustion process represented by the histograms in Figs. \ref{Fig_five}a and b 
can be identified as a  process with an ambiguous position of maximum. 
This effect was also found in the paper by Litak {\em et al.} \cite{Litak2005}.
In fact the double peak structure is clearly visible in Fig. \ref{Fig_five}b. 
They are directly related to
compression and combustion phenomena. 
Note that the exponential distribution in Fig. \ref{Fig_five}a indicates that the 
measurement process is not continuous, similar to the situation in a  cutting process with 
additional cracking effects \cite{Radons2004}. 
In our case, identification of  the pressure peak is not
certain because from time to time, especially for weak combustion 
it is interchanged with compression. On the basis of our experimental 
data we cannot  entirely  exclude the possibility of misfires 
\cite{Piernikarski2000}.
In Figs. \ref{Fig_five}c-d and Figs. \ref{Fig_five}e-f,
we show the distributions of $p_{max}$ and $\alpha_{pmax}$ for $\Delta \alpha_z=15^o$ and $\Delta \alpha_z=30^o$, respectively. Comparing these 
figures with
 Figs. \ref{Fig_five}a-b, we see that they are substantially different having more or less  the Gaussian 
shape with small asymmetry. Again we observe a characteristic broadening in $p_{max}$ and a narrowing in $\alpha_{pmax}$with increasing $\Delta \alpha_z$ .

\subsection{Multiscale Entropy}

Many physical systems evolve on multiple temporal and/or spatial scales and
are governed by complex dynamics. In recent years there has been a great deal of
interest in quantifying the complexity of these systems. However, no clear and
unambiguous definition of complexity has been established in the literature.
Intuitively, complexity is associated with "meaningful structural richness"    
\cite{Grassberger1991}. Several notions of entropy have been introduced to describe
complexity in a more precise way. But, as Costa et al.  \cite{Costa2002}  have pointed out,
these traditional entropy measures quantify the regularity (predictability) of the
underlying time series on a single scale, and there is no direct correspondence
between regularity and complexity. For instance, neither completely predictable
(e.g., periodic) signals, which have minimum entropy, nor completely unpredictable
(e.g., uncorrelated random) signals, which have maximum entropy, are truly complex.
For complex systems with multiple temporal or spatial scales, a definition of
complexity should include these multiscale features. Recently Costa et al.  \cite{Costa2002}
introduced the concept of multiscale entropy (MSE) to describe the complexity of  
such multiscale systems. They have successfully used this concept to describe the
nature of complexity in physiological time series such as those associated with
cardiac dynamics \cite{Costa2005} and gait mechanics \cite{Costa2003}. 
In this paper we perform a multiscale entropy analysis as a measure of complexity in
cycle-to-cycle variations of maximum pressure and peak pressure angle in a spark ignition engine. 
MSE is based on a coarse-graining procedure and can be carried out on a time series as
follows.

%f6
\begin{figure}
\vspace*{0.0cm}
\centerline{\includegraphics[width=9.0cm,angle=0]{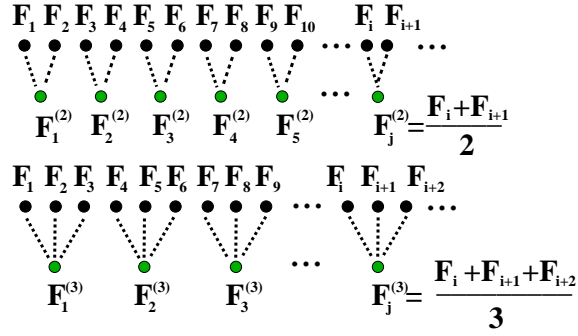}}
\caption{ \label{Fig_six} Schematic diagram of the coarse-graining procedure used in the calculation of
multiscale sample entropy
(after [34]). Note,
$F^{(1)}_i=F_i$ and $\tau$ is the scale factor.}
\end{figure}

For a given time series 
$\{F_1,F_2,...,F_n\}$, where $F_i=F(i)$, multiple coarse-grained time
series are constructed by
averaging the data points within non-overlapping windows of increasing length as  
shown in Fig. \ref{Fig_six}. Each element of the coarse-grained time series is computed
according to the equation:
\begin{equation}
F_j^{(\tau)}= \frac{1}{\tau} \sum_{i=(j-1) \tau +1}^{j \tau} F_i.
\end{equation}

Here  $\tau$ represents the scale factor, and  $ 1 \le j \le N/\tau$, the length of
each coarse-grained
time
series being equal to $N/\tau$. Note that for  $\tau = 1$, the coarse-grained time
series is simply
the original time series.

%f7
\begin{figure}
\vspace*{0.0cm}
\centerline{\includegraphics[width=5cm,angle=-90]{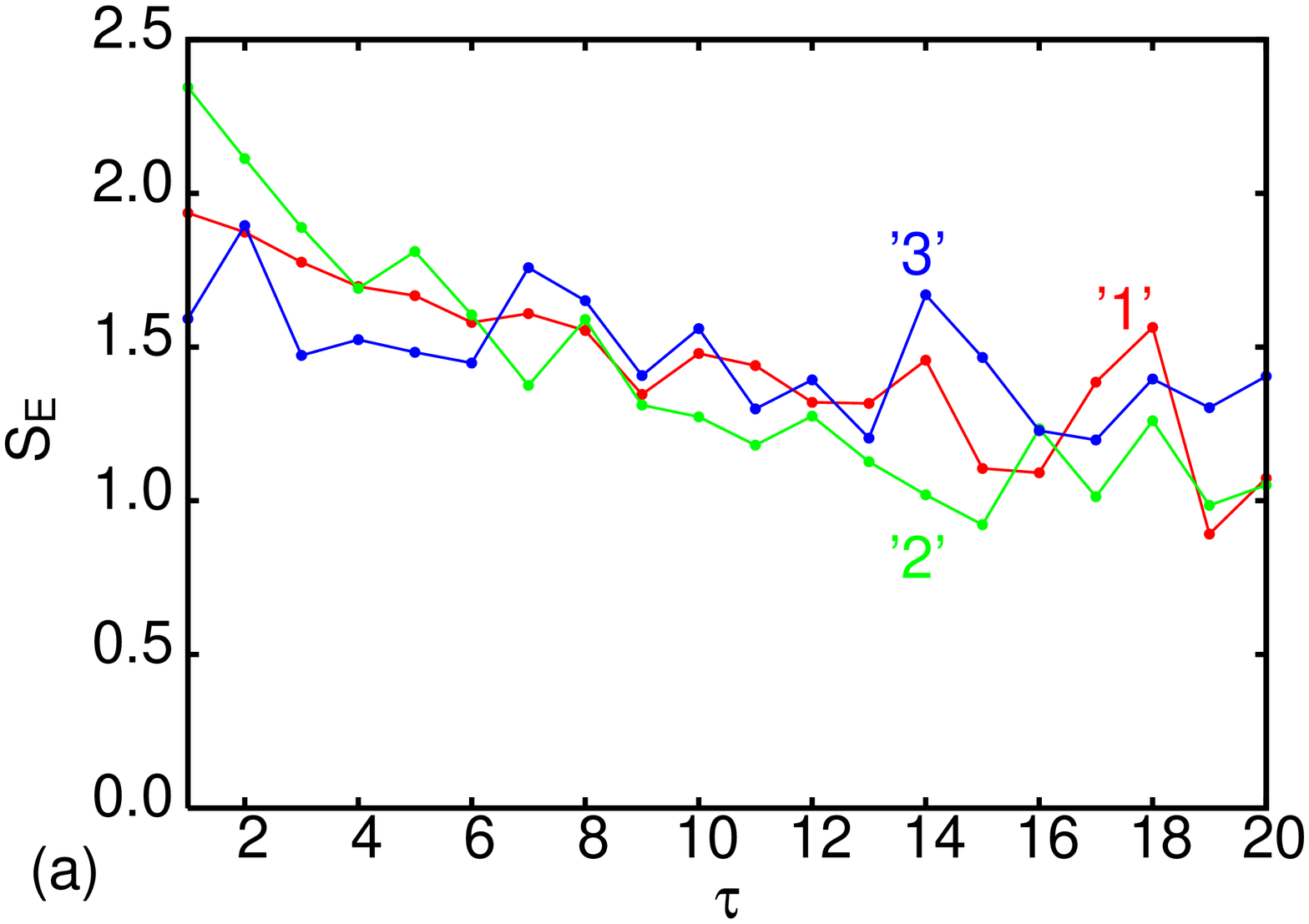} \hspace{-1cm}
\includegraphics[width=5cm,angle=-90]{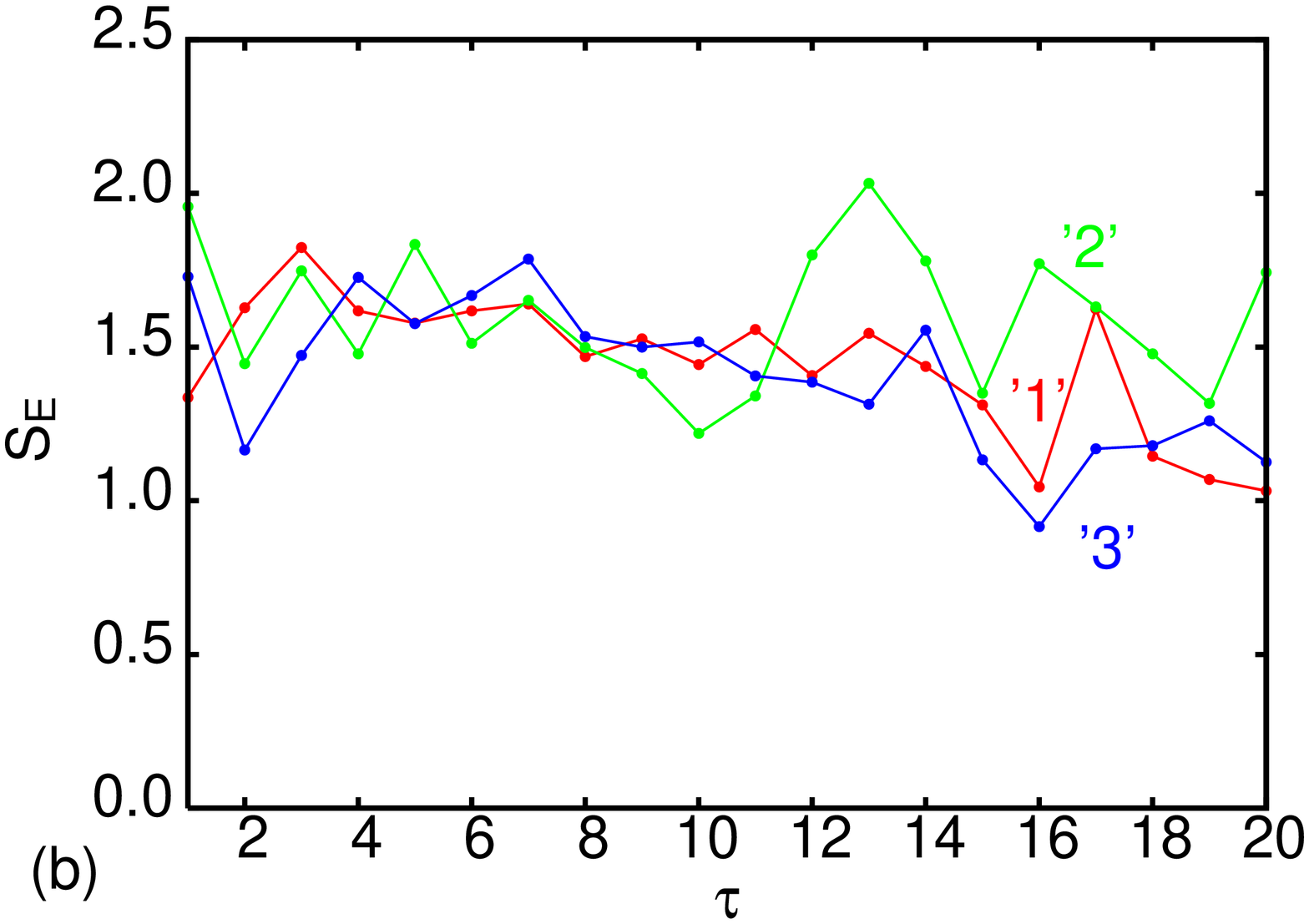}
}
\caption{\label{Fig_seven}
Sample entropy $S_E$ calculated for normalized time
series $F_i=p_{max}(i)/\sigma_(p_{max})$ (a) and
$\alpha_{pmax}/\sigma(\alpha{pmax})$ (b) versus the scale factor $\tau$.
$\sigma(p_{max})$ and
$\sigma(\alpha_{pmax})$ denote square root deviations of $p_{max}$ and
$\alpha_{pmax}$, respectively.
The length of chain $m=3$ and similarity factor
$r=0.15$ for all time series. Curves '1', '2' and '3'
correspond to $\Delta \alpha_z=5^o$,  15$^o$ and 30$^o$, respectively.
}
\end{figure}

The sample entropy $S_E$, proposed originally by 
Richman and  Moorman \cite{Richman2000}
  is calculated for the sequence  
of $m$ consecutive  data points which are
 'similar' to each other and will remain similar when one or more consecutive data 
points are
 included  \cite{Richman2000}. Here 'similar' means that the value of a specific measure of 
distance is
 less than a prescribed amount $r$. Therefore, the sample entropy depends on two parameters
$m$ and $r$. We have
 
\begin{equation}
 S_E(m,r,N)= \ln \frac{ \sum_{i=1}^{N-m} \{\tilde 
F(i,m,r)\}}{\sum_{i=1}^{N-m-1} \{\tilde F(i,m+1,r)\}}
\end{equation}

The sample entropy has been  calculated for each of the coarse-grained time 
series of $p_{max}$ and $\alpha_{pmax}$
and
then
plotted as a function of the scale factor $(\tau)$.
The results are shown in Fig. \ref{Fig_seven}, where
Fig. \ref{Fig_seven}a corresponds to  $p_{max}$ and Fig. \ref{Fig_seven}b to
$\alpha_{pmax}$.
It should be noted that for a simple but stochastic system the sample 
entropy $S_E$ has a hyperbolic shape. This shape can be identified only 
for curve '2' in Fig. \ref{Fig_seven}a while  the curves '1' and '3' are definitely more 
flat. In  Fig. \ref{Fig_seven}b,  on the other hand, all curves have a mild
dependence on $\tau$, and 
can be described approximately by horizontal lines. 
Such horizontal lines are typical for strongly asymmetric distributions 
such as 1/f noise.
In fact, if one looks more carefully at the slopes of 
the histograms of our signals, one can notice the obvious asymmetry in Figs. \ref{Fig_five}b and e, and 
to some extent
in  Figs. \ref{Fig_five}d,f.
Note that the peculiar nonlinear dynamics and a limited length of time series can additionally disturb 
these characteristics and produce fluctuations along the $\tau$ axis (Fig. 
\ref{Fig_seven}). 
    Interestingly the results show that
 $p_{max}$ and $\alpha_{pmax}$
are characterized by quite different aspects of the same dynamical 
process. 

\section{Discussion and Conclusions}

We have used the maximum pressure $p_{max}$ and the peak pressure angle $\alpha_{pmax}$
as variables for describing the combustion process in a spark ignition engine.
Their cycle-to-cycle variations are analyzed using different techniques such as return maps,   
histograms and multiscale entropy. 
Until now, researchers have used only one of these variables to formulate
an objective function  for optimization and control of engine dynamics.
Prompted by these complementary analyses of $p_{max}$ and
$\alpha_{pmax}$ presented in the literature
\cite{Litak2005,Eriksson1999,Nielsen1998},
we propose to use a mutual criterion for combustion efficiency
in
the SI engine. Instead of minimizing  the fluctuations of $p_{max}$ or $\alpha_{pmax}$
by looking at their smallest standard deviation $\sigma (p_{max})$, or
$\sigma (\alpha_{pmax})$:
%eq4.1
\begin{eqnarray}
&(a)~~~~~~& \sigma (p_{max}) = \frac{1}{\delta p_{max}}
\sqrt{  \frac{1}{N_{total}} \sum_{i=1}^{N_{total}}
\left(p_{max}(i)-\overline{p_{max}}\right)^2} \label{eq4.1} \\
&(b)~~~~~~& \sigma (\alpha_{pmax}) = \frac{1}{\delta
\alpha_{pmax}}
\sqrt{  \frac{1}{N_{total}} \sum_{i=1}^{N_{total}}
\left(\alpha_{pmax}(i)-\overline{\alpha_{pmax}}\right)^2} \nonumber
\end{eqnarray}
we look for a
two-variable square deviation $\sigma (p_{max},\alpha_{pmax})$ given by.
\begin{eqnarray}
&(c)~~~~~~& \sigma (p_{max},\alpha_{pmax}) = \nonumber \\
& &
\sqrt{  \frac{1}{N_{total}} \sum_{i=1}^{N_{total}}
\left( \frac{\left(p_{max}(i)-\overline{p_{max}}\right)^2}{(\delta
p_{max})^2} +
 \frac{ \left(\alpha_{pmax}(i)-\overline{\alpha_{pmax}}\right)^2}{(\delta
\alpha_{pmax})^2}
\right)}, \label{eq4.2}
\end{eqnarray}
where $\delta p_{max}$ and $\delta \alpha_{pmax}$
are chosen as the largest of $\sigma_{pmax}$ and $\sigma (\alpha_{pmax})$
from all examined cases and are treated as normalization.

%f8
\begin{figure}
\vspace*{0.0cm}
\centerline{\includegraphics[width=6cm,angle=-90]{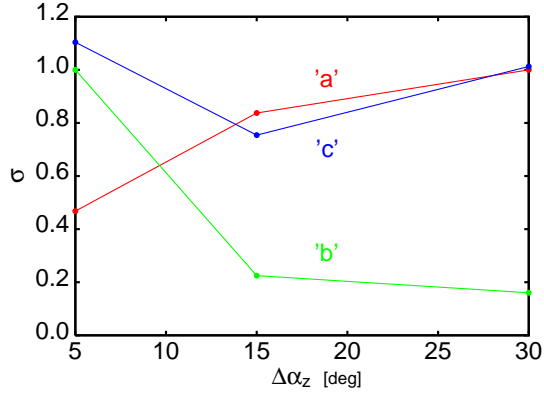}}   
\caption{\label{Fig_eight} Square deviations  $\sigma p_{max}$ (a),  $\sigma
(\alpha_{pmax})$ (b) and $\sigma (p_{max},\alpha_{pmax})$ (c) as a
generalized efficiency criterion (notation (a)-(c) as in Eqs. \ref{eq4.1}-\ref{eq4.2}).
}
\end{figure}

All the above square deviations are plotted in Fig. \ref{Fig_eight}. In this figure note that the plot for
$\sigma (p_{max},\alpha_{pmax})$ has a minimum when $\Delta \alpha_z=15^o$.
In our previous paper\cite{Kaminski2004}, where we analyzed noise effects 
in the cycle-to-cycle
fluctuations of  heat release, we noticed that  $\Delta \alpha_z=15^o$  
is  the case of optimal combustion conditions.
Now we see that for $\Delta \alpha_z=30^o$ the oscillations of  $p_{max}$ 
are the largest. 
On the other hand, for $\Delta \alpha_z=5^o$ the fluctuations in $\sigma (\alpha_{pmax})$ 
are significant.
The average heat release which varies with the spark advance angle has magnitudes of 352.54J for $\Delta \alpha_z=5^o$,
396.67J for $\Delta \alpha_z=15^o$ and  377.04J for $\Delta
\alpha_z=30^o$ 
\cite{Kaminski2004}.
It is the
largest  for  $\Delta \alpha_z=15^o$
indicating the largest burning rate of fuel while
the output torque, for the same speed of the crankshaft, is changing
from the largest value:
$S=30$Nm, in the case of $\Delta \alpha_z=30^o$, to a slightly smaller value:
 $S=28$Nm for $\Delta \alpha_z=15^o$, and to an even smaller value:
$S=21$Nm for  $\Delta \alpha_z=5^o$.
Knowing that the fresh fuel rate was the same in all the cases, we     
concluded that there were better combustion conditions for larger spark 
advance angles.
Additionally, in case of $\Delta \alpha_z=5^o$ some amount of fuel was not
burned because of weak combustion or misfires.  
In the previous paper  \cite{Kaminski2004} our analysis of noise level  led us 
to the conclusion that
in the case of $\Delta \alpha_z=15^o$, the combustion process was  most stable.

%f9
\begin{figure}  
\vspace*{0.0cm}

\centerline{ \includegraphics[width=4.5cm,angle=0]{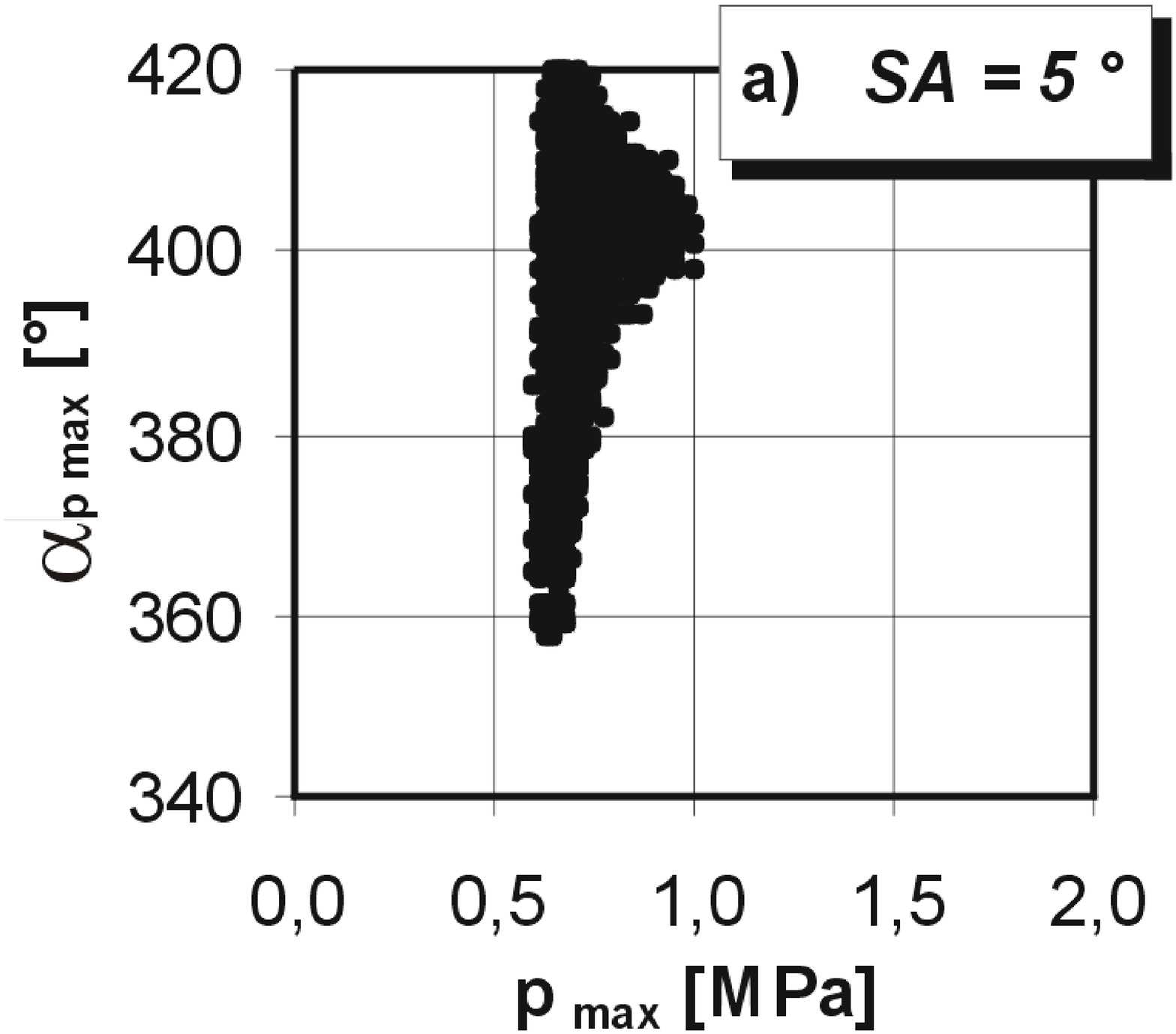}
\hspace{-0.3cm}  \includegraphics[width=4.5cm,angle=0]{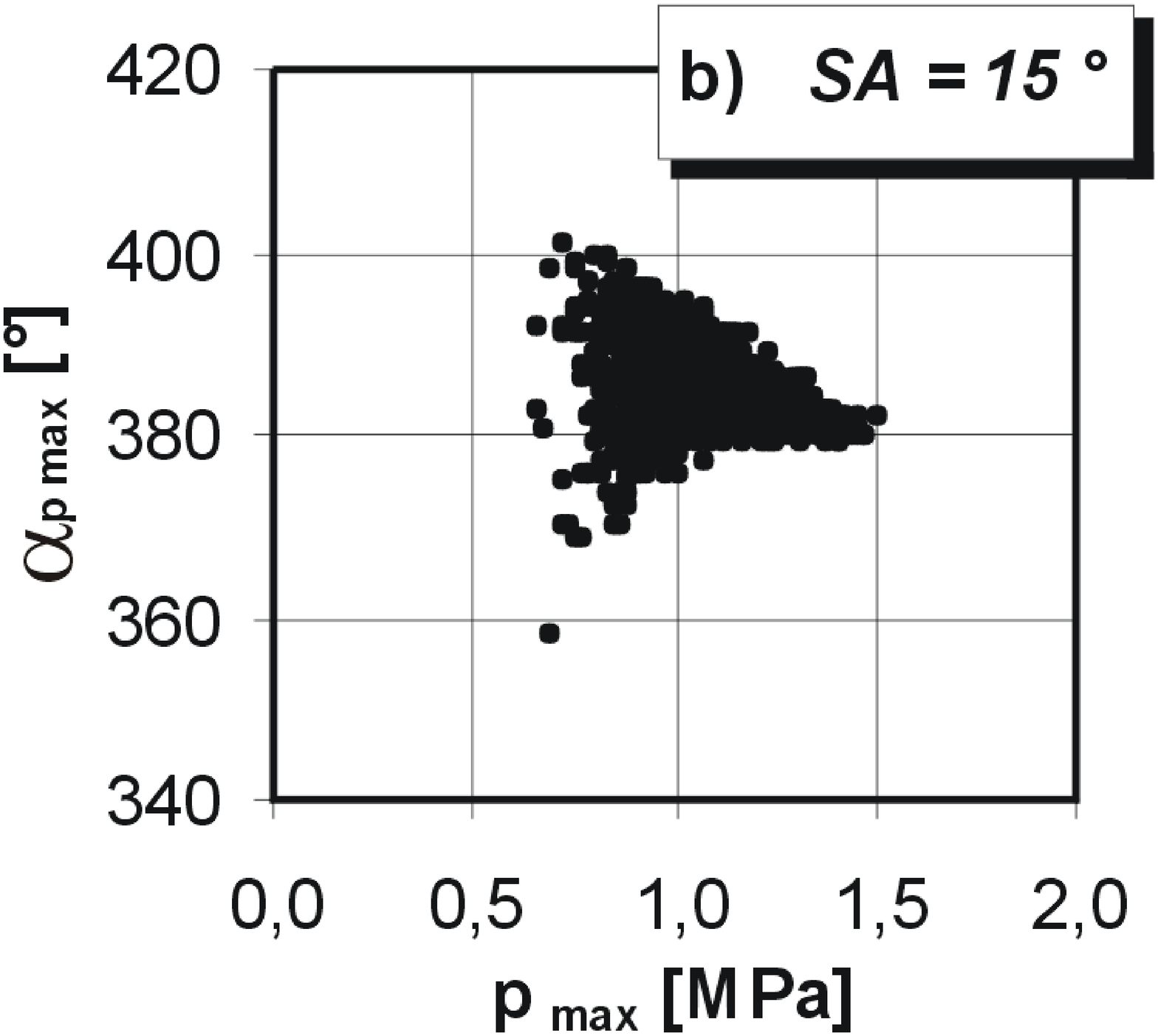} 
\hspace{-0.3cm}  \includegraphics[width=4.5cm,angle=0]{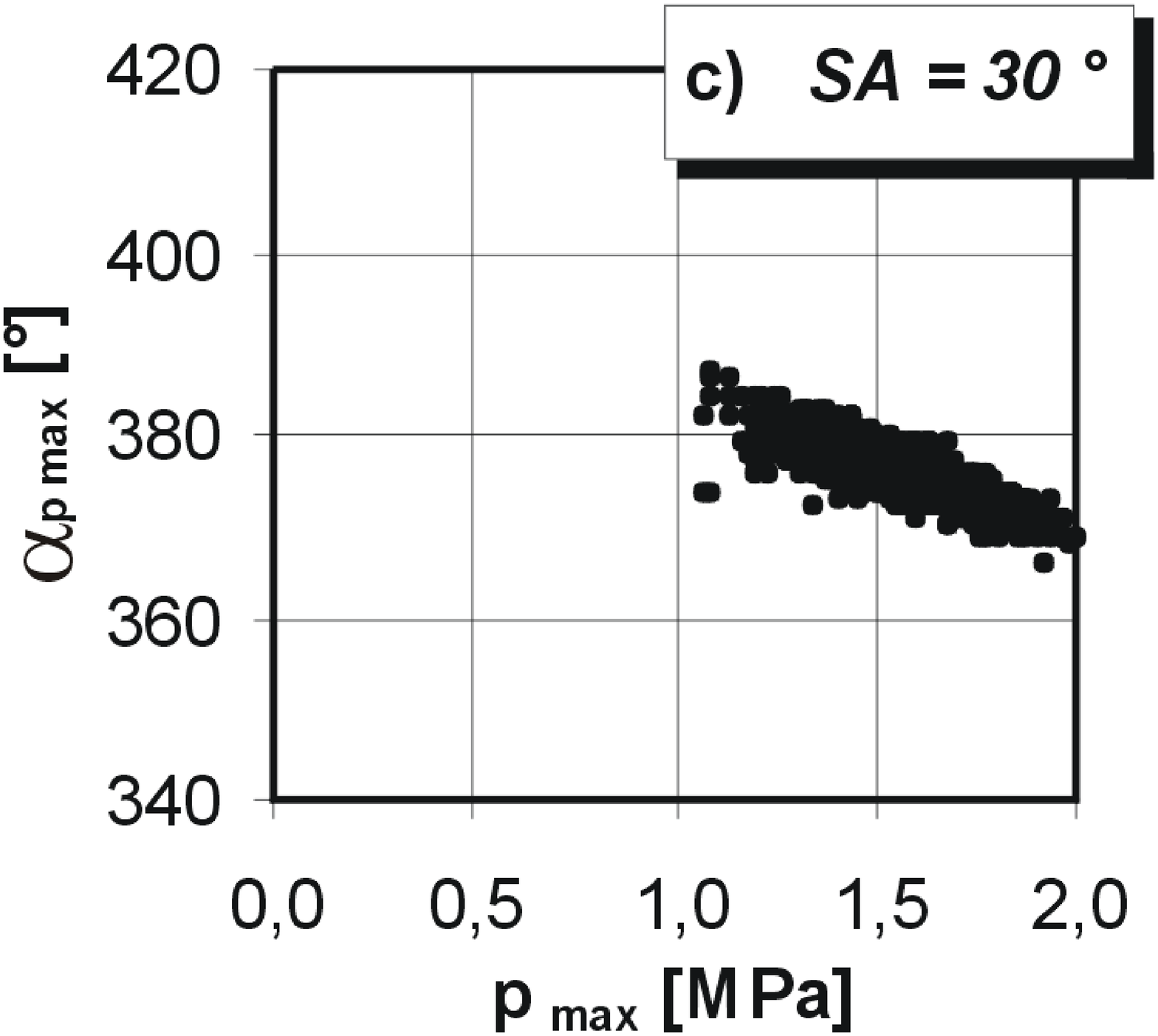}}
\vspace{-0.5cm} ~
(a) \hspace{3.8cm} (b) \hspace{3.8cm} (c)
\vspace{0.5cm}

\caption{\label{Fig_nine}
Maximum pressure-angle diagrams
($p_{max}$, $\alpha_{pmax}$) for three advance angles ($SA=\Delta \alpha_z=5^o$ (a), $15^o$
(b),~$30^o$ (c)).}
\end{figure}

Here we have found another argument leading to the same conclusion based  
on the two variables  $p_{max}$ and $\alpha_{pmax}$. Furthermore, from the 
multiscale entropy analysis we find that the $p_{max}$ sequence  is governed 
by the simplest dynamics. 
Examining both   $[p_{max}, \alpha_{pmax}]$ time series simultaneously 
it becomes clear that  the two variable square deviation $(\sigma (p_{max},\alpha_{pmax}))$
may play an 
important role 
 in the optimization procedure.
In Figs. \ref{Fig_nine}a-c we plotted 
the variables for each combustion cycle.
Following our philosophy, these plots  may be called  'a bifurcation
diagram'.
They show the transition from the case of  $\alpha_{pmax}$-dominated  fluctuations 
(Fig.  \ref{Fig_nine}a for $\Delta \alpha_z=5^o$) to the case of  $p_{max}$-dominated fluctuations 
(Fig.  \ref{Fig_nine}c for $\Delta \alpha_z=30^o$) through the intermediate case
 (Fig.  \ref{Fig_nine}b for $\Delta \alpha_z=15^o$).
One can easily see the qualitative change in the fluctuation type
starting from a vertical line structure in Fig. \ref{Fig_nine}a and ending with 
an almost horizontal line in 
Fig. \ref{Fig_nine}c.  
It is worth pointing out that both types of fluctuations are harmful for a stable 
engine operation and may result in decreased 
power output. It appears that the intermediate case is a 
compromise between these two types of fluctuations.
The detailed analysis of the relations between 
$p_{max}$ and $\alpha_{pmax}$ can provide 
further useful information on combustion dynamics as the rate of 
slow and/or fast individual  burning processes during engine use
corresponding to slow and fast heat release.
The two extreme realizations of a combustion process
presented in Fig.  \ref{Fig_nine}a and Fig.  \ref{Fig_nine}c (for $\Delta  
\alpha_z=5^o$ and  $30^o$, respectively) can be directly related to
slow and fast burning \cite{Ozdor1994}. 
Obviously the fast burning is not so effective as slow burning because it may
involve some events corresponding to a spatially limited engine volume.
Thus, as expected,  
if slow burning dominates, then the fluctuations of $p_{max}$ are small  (Figs. 
 \ref{Fig_nine}a and  \ref{Fig_three}a),
whereas  for fast burning the fluctuations of  $p_{max}$ are larger (Figs.    
 \ref{Fig_nine}c and  \ref{Fig_three}c).
Interestingly, measured fluctuations of $\alpha_{pmax}$ 
are 
governed mainly by the
 ambiguity of identification of maximum pressure  and possibly by a 
misfire phenomenon \cite{Piernikarski2000}. 
These two effects have   
opposite influences on 
$p_{max}$ and $\alpha_{pmax}$.   
On this basis our formulation of
optimization procedure (Eq. \ref{eq4.2}) seems to be a natural consequence
of a detailed analysis of the pressure-angle diagrams Figs. \ref{Fig_nine}a-c.
The minimum standard deviation defined in the 
$(p_{max},\alpha_{pmax})$
plane, 
 realized for $\Delta \alpha_z=15^o$, can be regarded as a compromise 
between the maximizing of power output and  
minimizing of pressure fluctuations. Clearly, larger power is accompanied by 
unstable combustion with larger pressure oscillations.

 It is worth noting that our simple mutual criterion seems to unify these 
approaches. 
Using this criterion we can associate the optimal combustion 
conditions with the spark advance angle $\Delta \alpha_z=15^o$. Of course, for any practical 
application we would need to consider many more values of $\Delta \alpha_z$ and more importantly 
check the output torque \cite{Heywood1988}.

\section*{Acknowledgements}
GL is grateful to Indiana University for hospitality.

%GL is grateful to Max Planck Institute for Physics of 
%Complex Systems in Dresden, where a part of 
%analysis has been done, for hospitality.

%\section*{Appendix A}
%\def\thesection{A}
%\def\theequation{A.\arabic{equation}}  % dla stylu 'article'

%\section*{Appendix B}
%\def\thesection{B}
%\def\theequation{B.\arabic{equation}}  % dla stylu 'article'


\begin{thebibliography}{99}

%1
\bibitem{Clerk1886} D. Clerk,  {\em The Gas Engine}, (Longmans, Green \& Co.,
London 1886).
  
%2
\bibitem{Patterson1966} D.J. Patterson,
Cylinder pressure variations, a fundamental combustion problem,
{\em SAE paper} No. 660129, 1966.

%3
\bibitem{Hubbard1976} M. Hubbard, P.D. Dobson, and J.D. Powell,
Closed loop control of spark advance using a cylinder pressure sensor,
{\em Journal of Dynamic Systems, Measurement and Control}  (1976) 414--420.

%4
\bibitem{Matekunas1986} F.A. Matekunas, Engine combustion control with
ignition timing by pressure ratio management, {\em US Pat}., A 4622939 Nov. 18
1986.

%5
\bibitem{Sawamoto1987} K. Sawamoto, Y. Kawamura, T. Kita and K.
Matsushita,
 Individual cylinder knock control by detecting cylinder pressure.
{\em SAE paper} No. 871911, 1987.

  
%6
\bibitem{Wagner1993}  R.M. Wagner, C.S.  Daw, and  J.F Thomas,
Controlling chaos
in spark-ignition
engines.
 in {\em Proceedings of the Central and Eastern States Joint Technical Meeting
of the
Combustion Institute} (New Orleans, 1993 March 15-17).

%7
\bibitem{Eriksson1997} L. Eriksson, L. Nilsen, M. Glavenius,
Development of control algorithm stabilizing torque for optimal position
of pressure peak,
{\em SAE Transactions Journal of Engines} {\bf 106} (1997) 1216--1223.
%{\em SAE paper} No. 970854, 1997.

  
%8 
\bibitem{Heywood1988} 
J.B. Heywood, {\em Internal Combustion Engine Fundamentals},
McGraw-Hill, New York 1988.

%9
\bibitem{Hu1996}
Z. Hu,
Nonlinear instabilities of combustion processes and cycle-to-cycle
variations in spark-ignition engines,
{\em SAE paper} No. 961197, 1996.

%10
\bibitem{Wagner2001}
R.M. Wagner, J.A. Drallmeier, and C.S. Daw,
Characterization of lean
combustion instability in pre-mixed charge spark ignition engines,
{\em International Journal of Engine Research} {\bf 1} (2001) 301--320.

%11
\bibitem{Daw2003}  
C.S. Daw, C.E.A. Finney, and E.R. Tracy,
A review of symbolic analysis of experimental time series,
{\em Rev. of Scen. Instr.} {\bf 74} ( 2003) 915--930.

%12
\bibitem{Wendeker2003}
M. Wendeker, J. Czarnigowski, G. Litak, and K. Szabelski,
Chaotic combustion in spark ignition engines,
{\em Chaos, Solitons \& Fractals} {\bf 18} (2003) 803--806.


%13
\bibitem{Kaminski2004}
T. Kami\'nski, M. Wendeker, K. Urbanowicz, and G. Litak,
Combustion process  in a spark ignition
engine: dynamics and noise level estimation,
{\em Chaos} {\bf 14} (2004) 461--466.

%14
\bibitem{Winsor1973} R.E. Winsor and  D.J. Patterson,
Mixture turbulence -- a key to cyclic combustion variation, 
{\em SAE paper}  No. 730086, 1973.


%15
\bibitem{Daily1988}
J.W. Daily,
"Cycle-to-cycle variations: a chaotic process?"
{\em Combustion
Science and
Technology} {\bf 57} (1988) 149--162.

  
%16
\bibitem{Kantor1984} J.C. Kantor,
A dynamical instability of spark-ignited
engines",
{\em Science} {\bf 224} (1984)   
1233--1235.

%17
\bibitem{Foakes1993} A.P. Foakes and D.G.  Pollard,
Investigation of a chaotic mechanism
for
cycle-to-cycle variations,
{\em Combustion Science and Technology} {\bf 90} (1993)
281--287.
   

%18
\bibitem{Chew1994}
L. Chew, R. Hoekstra, J.F. Nayfeh, and J. Navedo,
Chaos analysis of
in-cylinder pressure measurements,
{\em SAE paper}  No. 942486, 1994.


%19
\bibitem{Letellier1997} C. Letellier, S. Meunier-Guttin-Cluzel,
G. Gouesbet, F. Neveu, T. Duverger, and
B. Cousyn,
Use of the nonlinear dynamical system theory to study cycle-to-cycle
variations from spark-ignition engine pressure data,
{\em SAE paper}  No. 971640, 1997.

%20
\bibitem{Daw1996}  C.S. Daw, C.E.A. Finney, J.B. Green, Jr., M.B. Kennel,
J.F. Thomas, and F.T. Connolly,
A simple model for cyclic variations in a spark-ignition engine,
{\em SAE paper}  No. 962086, 1996.


   


%21
\bibitem{Daw1998} C.S. Daw, M.B. Kennel, C.E.A. Finney, and F.T. Connolly,
Observing  
and
modelling dynamics in an
internal combustion engine,
{\em Physical  Review} E {\bf 57} (1998)
2811--2819.

%22
\bibitem{Green1999}
J.B. Green Jr, C.S. Daw, J.S. Armfield, C.E.A. Finney, R.M. Wagner,
J.A. Drallmeier, M.B. Kennel, and
P. Durbetaki,
Time irreversibility and comparison of
cyclic-variability
models,
{\em SAE paper} No. 1999-01-0221, 1999.


   
%23
\bibitem{Wendeker2004}
M. Wendeker,
G. Litak, J. Czarnigowski, and K. Szabelski,
Nonperiodic oscillations in a spark ignition engine,
{\em Int. J. Bifurcation and Chaos} {\bf 14} (2004) 1801--1806.

%24
\bibitem{Nielsen1998}
L. Nielsen and L. Eriksson,
An ion-sense engine-fine-tuner,
{\em IEEE Control Systems Magazine} {\bf 18} (1998) 43­-52.
%25
\bibitem{Kaminski2005} T. Kaminski, PhD thesis, Technical University of Lublin, Lublin 2005.

%26
\bibitem{Matekunas1983} F.A. Matekunas,
Modes and measures of cyclic combustion variability,
{\em SAE paper}  No. 830337, 1983.
   
%27
\bibitem{Ozdor1994} N. Ozdor, M. Dulger, and E. Sher,
Cyclic variability in spark ignition engines:
a literature survey,
{\em SAE paper} No. paper 940987,
1994.

%28
\bibitem{Taccani2003}
R. Radu and R. Taccani,
Experimental setup for the cyclic variability analysis
on a spark ignition engine, {\em SAE NA paper} No. 2003-01-19 (2003).

%29
\bibitem{Litak2005}
G. Litak, R. Taccani, R. Radu, K. Urbanowicz, M. Wendeker, J.A.  Ho\l{}yst,
and A. Giadrossi,
Estimation of the noise level using coarse-grained entropy of
experimental time series
of internal pressure in a combustion engine,
{\em Chaos, Solitons \& Fractals},
{\bf 23} (2005) 1695-1701.



%30
\bibitem{Litak2005a}  G. Litak, M. Wendeker, M. Krupa, and J. Czarnigowski,
A numerical study of a simple stochastic/ deterministic model of
cycle-to-cycle combustion fluctuations in spark ignition engines,
{\it Journal of Vibration and Control} {\bf 11}  (2005)  371-379.
   
%31
\bibitem{Radons2004} G. Radons, R. Neugebauer (Eds.) {\em Nonlinear Dynamic Effects of
Production
Systems}, (Wiley-VCH,
Weinheim 2004).



%32
\bibitem{Piernikarski2000} D. Piernikarski and J. Hunicz,
Investigation of misfire nature using optical
combustion sensor in a SI automotive engine,
{\em SAE paper} No. 2000-02-0549 (2000).   

%33
\bibitem{Eriksson1999} L. Eriksson,
Spark advance for optimal efficiency,
{\em SAE paper} No. 99-01-0548, 1999.

%34
\bibitem{Grassberger1991}
P. Grassberger, in {\em Information Dynamics},  (Eds.) H. Atmanspacher and H.
Scheingraber, (Plenum Press, New York 1991).

%35
\bibitem{Costa2002}  
M. Costa, A. L. Goldberger, C.-K. Peng, Multiscale analysis of complex biological
signals, {\em Phys. Rev. Lett.} {\bf 89}, (2002) 068102.

%36
\bibitem{Costa2003}
M. Costa, C.-K. Peng, A. L. Goldberger,  Multiscale analysis of human gait
dynamics, {\em Physica} A {\bf 330}  (2003)  53-60.

%37
\bibitem{Costa2005}
M. Costa, A. L. Goldberger, C.-K. Peng, Multiscale analysis of biological signals,
{\em Phys. Rev.} E {\bf 89}  (2005)  021906.


%38
\bibitem{Richman2000} J.S. Richman and J.R. Moorman,
Physiological time-series analysis using approximate entropy and sample entropy,
{\em Am. J. Physiol} {\bf 278}  (2000)  H2039-H2049.

\end{thebibliography}
\end{document}